\documentclass[aps,prc,twocolumn,groupedaddress,nofootinbib]{revtex4-2}

\usepackage{amsmath,amssymb,multirow,graphicx,color,pifont}
\usepackage{hyperref}
\def\Vec#1{\boldsymbol{#1}}

\begin{document}


\title{
Proximity effects of vortices in
neutron $^3P_2$ superfluids in neutron stars: \\
Vortex core transitions and covalent bonding of 
vortex molecules
}


\author{Michikazu Kobayashi}
\affiliation{School of Engineering Science, Kochi University of Technology, Miyanoguchi 185, Tosayamada, Kami, Kochi 782-8502, Japan}

\author{Muneto Nitta}
\affiliation{Department of Physics, and Research and Education Center for Natural Sciences, Keio University, Hiyoshi 4-1-1, Yokohama, Kanagawa 223-8521, Japan}
\affiliation{International Institute for Sustainability with Knotted Chiral Meta Matter, Hiroshima University, 1-3-2 Kagamiyama, Higashi-Hiroshima, Hiroshima 739-8511, Japan}


\date{\today}

\begin{abstract}

Neutron $^3P_2$ superfluids consisting of 
neutron pairs with the total angular momentum $J=2$, 
spin-triplet, and $P$ wave are
 believed to be realized in 
neutron star cores. 
Within the Ginzburg-Landau theory 
 it was previously found 
that 
a singly quantized vortex 
is split into 
 two half-quantized non-Abelian vortices
 connected by one (or three) soliton(s) 
 forming a vortex molecule with the soliton bond(s), 
 in the absence (presence) of a magnetic field parallel to them.
In this paper, we investigate 
proximity effects of two vortex molecules 
by exhausting all possible two vortex molecule states 
consisting of four half-quantized vortices 
and 
determine
the phase diagram 
spanned by 
the magnetic field and rotation speed.
As the rotation speed is increased, 
the distance between the two vortex molecules 
becomes shorter.
In the magnetic field below the critical value, 
we find 
that as the rotation speed is increased, 
the two separated vortex molecules transit 
to a dimerized vortex molecule, 
where the two vortex molecules   
are bridged by two solitons 
that we call  ``covalent bonds'' in analogy with chemical molecules.
We also find that 
the orders of the constituent half-quantized vortex cores transit 
from a ferromagnetic order to 
a cyclic order as the vortex molecules come closer.  
On the other hand, no dimerization occurs 
in the magnetic field above the critical value. 
Instead, we find a transition 
for the polarization direction of the vortex molecules from 
a configuration parallel to the separation 
to one perpendicular to the separation 
as they come closer.
We also show some examples of three, four, and many vortex molecule states.

\end{abstract}


\maketitle

\section{Introduction}

Neutron stars are rapidly rotating 
extremely high density compact stars 
accompanied with strong magnetic fields.
Recently, 
there have been great 
progresses in astrophysical observations 
of neutron stars (pulsars), such as
massive neutron stars with masses about twice as large as the solar mass~\cite{Demorest:2010bx,Antoniadis1233232}, 
detection of gravitational waves from a binary neutron star merger~\cite{TheLIGOScientific:2017qsa,Abbott:2020uma}, 
and 
the Neutron star Interior Composition Explorer (NICER) mission~\cite{Riley:2019yda,  Miller:2019cac}. 
These are providing us 
 astrophysical laboratories for exploring 
 nuclear and quantum chromodynamics (QCD) matter under extreme conditions: 
 extremely high density,  with strong magnetic fields and 
under rapid rotations 
\cite{Graber:2016imq,Baym:2017whm}.  

It is believed that 
the interior of neutron stars 
exhibit neutron superfluidity and proton superconductivity 
\cite{Migdal:1960}, see 
Refs.~\cite{Chamel:2008ca,Chamel2017,Haskell:2017lkl,Sedrakian:2018ydt,Graber:2016imq,Andersson2021} 
as a review.
Such supercomponents
provide low-energy excitations  
affecting several processes and properties 
of neutron stars: 
neutrino emissivities and specific heats relevant to the long relaxation time after pulsar glitches (sudden speed-up events 
of neutron stars)~\cite{Baym1969,Pines1972,Takatsuka:1988kx}, 
and the enhancement of neutrino emission around 
the critical point of the superfluid transition~\cite{Yakovlev:2000jp,Potekhin:2015qsa,Yakovlev:1999sk,Heinke:2010cr,Shternin2011,Page:2010aw}.  
In addition to these, 
vortices have quantized circulations
due to the Feynman-Onsager's quantization 
in order for the wave function to be single-valued,
and thus are called quantum vortices.
As a result, 
rotating superfluids  possess a large number of quantum vortices along the rotation axis, forming a vortex lattice. 
In typical neutron stars, 
there should exist $10^{17}$ quantum vortices, 
considered to play significant roles in neutron star dynamics.
For instance, the origin of pulsar glitches was suggested to be explained 
by avalanche unpinning of a large number of quantum  vortices~\cite{reichley,Anderson:1975zze}.

At lower density corresponding to outer cores of neutron stars,
Cooper pairs of neutrons responsible for neutron superfluidity 
are realized due to the attraction 
by the $^1S_0$ channel 
between two neutrons  
\cite{1966ApJ...145..834W}.
On the other hand, at higher density corresponding to 
the inner cores of neutron stars, 
the $^1S_0$ channel becomes repulsive 
due to the strong short-range repulsion. 
Instead, the $^3P_2$ channel  
 originating from a strong spin-orbit force at large scattering energy 
becomes more dominant, 
where 
neutron Cooper pairs possess a spin-triplet and $P$ wave 
with 
the total angular momentum $J=2$ 
\cite{Tabakin:1968zz,Hoffberg:1970vqj,Tamagaki1970,Takatsuka1971,Takatsuka1972,Fujita1972,Richardson:1972xn,Amundsen:1984qc,Takatsuka:1992ga,Baldo:1992kzz,Elgaroy:1996hp,Khodel:1998hn,Baldo:1998ca,Khodel:2000qw,Zverev:2003ak,Maurizio:2014qsa,Bogner:2009bt,Srinivas:2016kir}. 
Furthermore,  
 the $^3P_2$ channel is tolerant against the strong magnetic field  
such as $10^{15}$--$10^{18}$ G for magnetars, 
because aligned Cooper pairs with the spin-triplet pairing are not broken by the Zeeman effect,  
in contrast to the $S$-wave Cooper pairs which can survive at most 
around the magnetic field $10^{17}$ G \cite{Stein:2016}. 
In astrophysical observations, 
the rapid cooling of the neutron star in Cassiopeia A 
was proposed to be explained by the enhancement of neutrino emissivities due to the formation and dissociation of neutron $^3P_2$ Cooper pairs~\cite{Heinke:2010cr,Shternin2011,Page:2010aw}.

Theoretically, there are two frameworks to deal with 
 the $^3P_2$ superfluids related to each other: 
 a microscopic theory known as  
the Bogoliubov--de Gennes (BdG) equation 
describing fermion degrees of freedom, 
and 
the Ginzburg-Landau (GL) theory 
for the order parameters 
conveniently describing bosonic excitations. 
The latter can be obtained from the former 
by integrating out fermion degrees of freedom 
as an expansion of the order parameters and spatial derivatives. 
Thus, the GL theory is the low-energy effective theory 
describing large distance behaviors,   
which is valid only in the region close to the critical temperature.  
Among 
superfluid states with $J=2$ classified into nematic, cyclic, and ferromagnetic phases etc~\cite{merminPRA74}, 
the GL theory for $^3P_2$ superfluids 
 \cite{Fujita1972,Richardson:1972xn,
Sauls:1978lna,Muzikar:1980as,Sauls:1982ie,Vulovic:1984kc,Masuda:2015jka,Masuda:2016vak, 
Yasui:2018tcr,Yasui:2019tgc,Yasui:2019unp} 
predicts that the ground state is in 
the nematic phase 
at least in the weak coupling limit~\cite{
Sauls:1978lna,Muzikar:1980as,Sauls:1982ie} 
\footnote{
Among $J=2$ superfluids, 
nematic phases are also known in 
spin-2 Bose-Einstein condensates (BECs) 
of ultracold atomic gases 
\cite{Zhou:2006fd,Semenoff:2006vv,
Uchino:2010,Uchino:2010pf,Borgh:2016cco,Kobayashi:2021arv}, 
and thus they have common bosonic properties. 
}.
The nematic phase consists of almost degenerate 
three different states 
with different unbroken symmetries: 
uniaxial nematic (UN), $D_2$ biaxial nematic ($D_2$BN), 
and $D_4$ biaxial nematic ($D_4$BN) phases 
with unbroken groups $O(2)$, $D_2$ and $D_4$, respectively.
Here, $D_n$ is a dihedral group of order $n$ 
[see Table~\ref{table:manifold}(a) and (b)].
Depending on the magnetic field and temperature, 
the UN, $D_2$BN, or $D_4$BN state 
is realized as the ground state  
for zero magnetic field, nonzero one below the critical value $B_c$, 
or nonzero one above $B_c$, respectively 
\cite{Masuda:2015jka,Yasui:2018tcr,Yasui:2019tgc,
Yasui:2019unp} 
[see Table~\ref{table:manifold}(a)]. 
Apart from nematic phases,
more general uniform states (which do not have to be realized as 
the ground states) of $^3P_2$ superfluids were classified 
according to symmetries \cite{Kobayashi:2021arv}.
In fact, 
beyond the quasiclassical approximation,  
the ferromagnetic phase 
was found 
in the region close to the critical temperature 
\cite{Mizushima:2021qrz}. 
The GL approach is useful 
not only to determine the ground states 
but also to deal with 
bosonic collective excitations 
\cite{Bedaque:2003wj,Leinson:2011wf,Leinson:2012pn,Leinson:2013si,Bedaque:2012bs,Bedaque:2013rya,Bedaque:2013fja,Bedaque:2014zta,Leinson:2009nu,Leinson:2010yf,Leinson:2010pk,Leinson:2010ru,Leinson:2011jr} 
relevant for the cooling process of neutron stars,   
and
various topological excitations and defects, such as 
vortices (as explained below in more detail), 
domain walls \cite{Yasui:2019vci}, 
and 
the boundary defect (boojums) \cite{Yasui:2019pgb}. 

On the other hand,
the BdG approach 
offers a microscopic description 
with fermion degrees of freedom 
valid at short distances and all ranges of temperatures 
including zero temperature 
\cite{Mizushima:2016fbn,
Mizushima:2019spl,
Mizushima:2021qrz,
Masaki:2019rsz,Masaki:2021hmk}. 
It was applied to the phase diagram 
of $^3P_2$ superfluids
in the plane of 
the temperature and magnetic field, 
which is valid even at zero temperature  
\cite{Mizushima:2016fbn},
including a tricritical point connecting 
first and second order phase transition lines 
 between $D_4$ and $D_2$BN phases 
\cite{Mizushima:2016fbn,Mizushima:2019spl}. 
Furthermore, $^3P_2$ superfluids 
were shown to be topological superfluids 
of a class DIII in the classification of topological insulators and superconductors~\cite{Schnyder:2008tya,Ryu:2010zza}, 
ensuring a topologically protected 
gapless Majorana fermion on its boundary 
\cite{Mizushima:2016fbn} 
and inside vortex cores 
\cite{Masaki:2019rsz,Masaki:2021hmk} 
as explained below.

Since it is promising that quantum vortices 
in $S$-wave superfluids play significant roles 
in neutron star dynamics, the same should be expected 
for the $^3P_2$ superfluids as well.  
 In fact, 
quantum vortices were investigated 
in the case of $^3P_2$ superfluids 
both in the GL theory 
\cite{Muzikar:1980as,Sauls:1982ie,Fujita1972,Masuda:2015jka,Chatterjee:2016gpm,Masuda:2016vak,Kobayashi:2022moc} 
(coreless vortices \cite{Leinson:2020xjz})  
and in the BdG theory 
\cite{Masaki:2019rsz,Masaki:2021hmk}.
The first homotopy group classifies types of vortices 
in each phase \cite{Masuda:2015jka} 
as in Table \ref{table:manifold}(d).
Singly quantized vortices in $^3P_2$ superfluids 
were studied in the GL theory \cite{Muzikar:1980as,Sauls:1982ie,Fujita1972,Masuda:2015jka,Chatterjee:2016gpm} 
and in the BdG theory \cite{Masaki:2019rsz} 
with topologically protected Majorana fermions in 
the vortex cores.
Vortices more peculiar to 
the $^3P_2$ superfluids 
are half-quantized non-Abelian vortices
\cite{Masuda:2016vak,Masaki:2021hmk,Kobayashi:2022moc}, 
which have 
a half of the Feynman-Onsager's quantized circulations  
and are characterized by 
 a non-Abelian first homotopy group $D_4^*$, thus giving noncommutativity  
when exchanging two vortices. 
Isolated half-quantized non-Abelian vortices 
are topologically allowed only in the $D_4$BN phase. 
The existence of half-quantized vortices 
was proposed to explain a scaling law of 
pulsar glitches \cite{Marmorini:2020zfp}.
\begin{table}
\caption{
\label{table:manifold}
Properties of $^3 P_2$ superfluids 
in the absence and presence of the magnetic field 
below and above the critical magnetic field $B_c$. 
(a) Phases in the bulk, (b) unbroken symmetries in the bulk, 
(c) OPMs corresponding to the symmetry breakings, 
(d) the first homotopy groups 
of the OPMs supporting vortices, 
(e) the order inside cores of half-quantized vortices, 
(f) the number of solitons connecting 
two half-quantized vortices consisting of a singly quantized vortex, 
and 
(g) the order of soliton cores are summarized.
In row (d), 
${\mathbb Q} =  D_2^\ast $ is a quaternion group 
as the universal covering group of $D_2$
with the astarisk $\ast$ denoting the universal covering, 
and 
 $\rtimes_h$ is a product defined in Ref.~\cite{Kobayashi:2011xb},
 supporting the 
 isolated half-quantized non-Abelian vortices in the $D_4$BN phase.
The brackets in rows (e) and (f) and the column $|\Vec{B}|=0$
represent a metastable solution.
The most parts of this table are taken from our previous paper \cite{Kobayashi:2022moc}.
}
\begin{tabular}{lccc} \hline\hline
 &  $|\Vec{B}| =0$ &  $0< |\Vec{B}| <B_c$ & $B_c < |\Vec{B}| $\\ \hline 
(a) Phase & UN & $D_2$BN & $D_4$BN \\
(b) Symmetry & $\mathrm{O}(2)$ & $D_2$ & $D_4$ \\
(c) OPM &
 $S^1 \times {\mathbb R}P^2$ 
& $\mathrm{U}(1) \times \frac{\mathrm{SO}(3)}{D_2}$ 
& $\frac{\mathrm{U}(1) \times \mathrm{SO}(3)}{D_4}$\\
(d) $\pi_1$(OPM) 
& ${\mathbb Z} \oplus {\mathbb Z}_2$ 
& ${\mathbb Z} \oplus {\mathbb Q}$ 
& ${\mathbb Z} \rtimes_h D_4^\ast $\\ 
\multirow{2}{*}{(e) Vortex core order} & Ferro & \multirow{2}{*}{Cyclic} & \multirow{2}{*}{Cyclic} \\[-3pt]
 & (Cyclic) & & \\
(f) $\#$ of solitons & 1 (3) & 3 & 3\\
(g) Soliton core order &  $D_4$BN & $D_4$BN & $D_2$BN \\ \hline\hline
\end{tabular}
\end{table}
While an axisymmetric ansatz was employed
in the previous studies of vortex solutions 
\cite{Muzikar:1980as,Sauls:1982ie,Fujita1972,Masuda:2015jka,Chatterjee:2016gpm,Masuda:2016vak}, 
it was shown 
in the BdG equation \cite{Masaki:2021hmk} 
that 
a singly quantized vortex always splits into two half-quantized non-Abelian vortices
with any strength of the magnetic field, 
forming a vortex molecule \footnote{
Similar molecules of half-quantized vortices 
connected by a linear soliton 
are present in the other multicomponent 
condensed matter physics and high energy physics: 
multicomponent or multigap superconductors 
\cite{Babaev:2001hv,Babaev:2004rm,Goryo2007,Tanaka2007,Crisan2007,Guikema2008,Nitta:2010yf,Garaud:2011zk,Garaud:2012pn,Garaud2012a,Tanaka2017,Tanaka2018,Chatterjee:2019jez,Garaud:2022izd},
coherently coupled multicomponent BECs \cite{Son:2001td,Mueller2002,Kasamatsu2003,Kasamatsu:2004tvg,Kuopanportti2012,Aftalion2012,Eto:2012rc,Cipriani:2013nya,Eto:2013spa,Nitta:2013eaa,Dantas2015,Tylutki:2016mgy,Eto:2017rfr,Eto:2019uhe,Kobayashi:2018ezm,MenciaUranga2018},
high density QCD \cite{Eto:2021nle}, 
and two-Higgs doublet models \cite{Eto:2021dca}.
However, 
the unique feature of $^3P_2$ superfluids is that 
constituent half-quantized vortices are 
{\it non-Abelian} vortices 
associated with 
a non-Abelian first homotopy group.
}.
It was also found in Ref.~\cite{Masaki:2021hmk} 
that a Majorana fermion zero mode is trapped in 
each half-quantized vortex. 
Such a splitting of a singly quantized vortex was also confirmed 
 in the GL theory without enforcing axisymmetry 
 \cite{Kobayashi:2022moc}.
In the GL theory,
cores of two half-quantized vortices 
exhibit a ferromagnetic order
in the UN phase with the zero magnetic field, 
and a cyclic order in the $D_2$ and $D_4$BN phases 
in the presence of the magnetic field,  
as summarized in Table \ref{table:manifold}(e).
In the UN phase in the absence of the magnetic field, 
the most stable singly quantized vortex configuration 
consists of
two half-quantized vortices with the ferromagnetic cores connected by 
a single soliton of the $D_4$BN order.
In addition to this, there is also a metastable configuration 
consisting 
of those of the cyclic cores 
 connected by three solitons of 
the $D_4$BN order.
On the other hand, 
in the $D_2$ ($D_4$)BN phase 
in the presence of small (large) magnetic field, 
two half-quantized vortices 
are connected by three linear solitons of 
the $D_4$ ($D_2$)BN order. 
[See Table \ref{table:manifold}(f) and (g).]
Even in the bulk UN and $D_2$BN phases,
the $D_4$BN order locally appears around the vortex cores as solitons,  
because isolated half-quantized vortices can topologically exist 
only in the $D_4$BN state and thus 
splitting into two half-quantized vortices is possible 
only inside the $D_4$BN order.

In this paper, we investigate 
proximity effects of two vortex molecules.
As explained above, a singly quantized vortex 
 is of the form of a vortex molecule of 
 two half-quantized non-Abelian vortices 
connected by a single soliton (three solitons) 
in the absence (presence) of a magnetic field.
In the absence of magnetic field, we find that 
a transition of half-quantized non-Abelian vortex cores occurs from a ferromagnetic order (connected by a single soliton)
to a cyclic order (connected by three solitons), 
when two vortex molecules come close to each other with increasing the rotation speed.
Furthermore, one of three solitons reconnects to one of the other vortex molecule and bridges the two vortex molecules, 
forming a ``covalent bond.'' 
Thus, the four constituent half-quantized vortices are connected 
by one or two soliton(s) alternately. 
In the presence of a magnetic field below the critical magnetic field $B<B_c$,
we find the same transition from two isolated vortex molecules to a ``dimerized molecule'' with a covalent bonding of solitons
when the two vortex molecules come close to each other with increasing 
the rotation speed \footnote{
A similar reconnection of solitons 
was observed in numerical simulations of two-component BECs 
\cite{Cipriani:2013nya,Eto:2019uhe},
and similar chemical bonds of vortex molecules were found 
in multicomponent superconductors 
with arbitrary charges and Josephson couplings 
\cite{Chatterjee:2019jez}.
}.
However, 
we do not find such phenomena 
in the $D_4$BN phase above the critical magnetic field.
Instead, we find another type of a transition for the polarization direction of the vortex molecules from one parallel to the separation 
of the two vortex molecules 
to the other perpendicular to the separation 
of the two vortex molecules 
with increasing the external rotation.
Then, we further study three and four vortex molecules.
In the $D_2$BN phase below the critical magnetic field,
as in the case of two vortex molecules, we find a transition from isolated vortex molecules to trimerized and tetramerized vortex molecules
consisting of six and eight half-quantized non-Abelian vortices.
In the end of the paper, we also discuss states with many vortex molecules as candidates for neutron star interiors.
Unlike singlet-paring superfluids, vortex configuration becomes irregular due to polymerization of vortex molecules or a frustration between spatial
configuration and the alignment of vortex molecules.


This paper is organized as follows.
In Sec.~\ref{sec:formulation}, we begin with  
formulations of $^3P_2$ superfluids within the GL 
approach in our notation, and shortly summarize our previous results for a single vortex molecule state.
In Sec.~\ref{sec:multivortex}, we show our main results for two, three, four, and many vortex molecule states.
Section~\ref{sec:summary} is devoted to a summary 
and discussion. 





\section{Ginzburg-Landau free energy and single-vortex molecule state for 
$^3P_2$ neutron superfluids}\label{sec:formulation}

We start from a brief review of the GL theory for $^3P_2$
 superfluids \cite{Yasui:2019unp} 
 reformulated in the notation of Ref.~\cite{Kobayashi:2021arv} and single-vortex molecule state within the GL formalism.
The details were discussed in Refs.~\cite{Yasui:2019unp,Kobayashi:2021arv} for GL theory and Ref.~\cite{Kobayashi:2022moc}
 for single-vortex state.

\subsection{Ginzburg-Landau theory}\label{subsec:GL-theory}

The effective GL Lagrangian density $f$ is given by 
\begin{align}
\begin{aligned}
f &= K_0 \left(f_{202}^{(0)} + f_{202}^{(1)}\right) + \alpha f_{002} + \beta_0 f_{004} + \gamma_0 f_{006} \\
&+ \delta_0 f_{008} + \beta_2 f_{022} + \gamma_2 f_{024} \\
&+ \sum_{4 l + 2 m + n = 10} \mathcal{O}(\nabla^l |\Vec{B}|^m A^n),
\end{aligned}
\label{eq:Lagrangian}
\end{align}
where $f_{lmn}$ is the free energy part including 
$l$ spatial derivatives $\nabla$, 
$m$th order of the magnetic field $\Vec{B}$, and $n$th order of spin-2 spinor order parameter $\psi = (\psi_2, \psi_1, \psi_0, \psi_{-1}, \psi_{-2})^T$.
The spatial derivative term $f_{202}$ is further separated into current-spin independent and dependent parts $f_{202}^{(0)}$ and $f_{202}^{(1)}$, respectively.
Each term can be written as
\begin{widetext}
\begin{align}
\begin{aligned}
& 
f_{202}^{(0)} = 3 \Vec{j}^\dagger \cdot \Vec{j}, \quad
f_{202}^{(1)} = 4 \Vec{j}^\dagger \cdot \Vec{j} - \frac{i}{2} \Vec{j}^\dagger \cdot \hat{\Vec{S}} \times \Vec{j} - \left( \Vec{j}^\dagger \cdot \hat{\Vec{S}} \right) \left( \hat{\Vec{S}} \cdot \Vec{j} \right), 
  \\
& f_{002} = 3 \rho, \quad
f_{004} = 6 \rho^2 + \frac{3}{4} \Vec{S}^2 - \frac{3}{2} |\Psi_{20}|^2, \\
& f_{022} = 2 \rho \Vec{B}^2 - \frac{1}{2} \psi^\dagger \hat{S}_{\Vec{B}} \hat{S}_{\Vec{B}} \psi, \quad
f_{006} = - 324 \rho^3 - 81 \rho \Vec{S}^2 + 162 \rho |\Psi_{20}|^2 + 15 |\Psi_{30}|^2 - 27 |\Phi_{30}|^2, \\
& f_{024} = \left( -106 \rho^2 + \frac{9}{2} \Vec{S}^2 + 31 |\Psi_{20}|^2 \right) \Vec{B}^2 \\
&\phantom{f_{024}} + \left( 22 \rho \psi^\dagger \hat{S}_{\Vec{B}} \hat{S}_{\Vec{B}} \psi + \mathrm{Re}\left[ \Psi_{20}^\ast \psi^T \hat{S}_{\Vec{B}}^T \hat{T} \hat{S}_{\Vec{B}} \psi \right] \vphantom{\frac{5}{4}} + \frac{5}{4} \Psi_{22}^\dagger \hat{S}_{\Vec{B}} \hat{S}_{\Vec{B}} \Psi_{22} + \frac{1}{2} \Phi_{22}^T \hat{S}_{\Vec{B}}^T \hat{T} \hat{S}_{\Vec{B}} \Phi_{22} \right), \\
& f_{008} = 6480 \rho^4 + 1944 \rho^2 \Vec{S}^2 - 5184 \rho^2 |\Psi_{20}|^2 - 864 \rho |\Psi_{30}|^2 + 2592 \rho |\Phi_{30}|^2 + 81 \Vec{S}^4 + 648 |\Psi_{20}|^4 - 1296 \Gamma_{4} .
\end{aligned}
\end{align}
\end{widetext}
Here, $\hat{S}_i$ ($i = x,y,z$) are $5 \times 5$ spin-2 matrices,
\begin{align}
\begin{aligned}
& \Vec{j} \equiv - i \nabla \psi, \quad
\end{aligned}
\end{align}
and the invariants are given by
\begin{align}
\begin{aligned}
& \rho \equiv \psi^\dagger \psi, \quad
\Vec{S} \equiv \psi^\dagger \hat{\Vec{S}} \psi, \\ 
& 
\Gamma_4 \equiv \mathrm{Re}\left[\Psi_{20} \Phi_{30}^{\ast 2}\right], \quad
\Psi_{20} \equiv \sqrt{5} 
C^{00}_{2 m_1,2 m_2} \psi_{m_1} \psi_{m_2}, \\
& \Psi_{30}
\equiv - \sqrt{\frac{35}{2}} 
C^{00}_{JM,2m_3} C^{JM}_{2m_1,2m_2} \psi_{m_1} \psi_{m_2} \psi_{m_3}, 
\\ 
& 
\Phi_{30}
\equiv - \sqrt{\frac{35}{2}} 
C^{00}_{JM,2m_3} C^{JM}_{2m_1,2m_2} \psi_{m_1} \psi_{m_2} \psi^\ast_{-m_3} \\
&\phantom{\Phi_{30}} \times (-1)^{m_3},
\end{aligned}
\end{align}
where we have taken the Einstein summation notation for $-2 \leq m_{1,2,3} \leq 2$, $0 \leq J \leq 4$, and $-J \leq M \leq J$ 
for $\Psi_{20}$, $\Psi_{30}$, and $\Phi_{30}$
with the Clebsch-Gordan coefficients $C^{JM}_{s_1m_1,s_2m_2}$.

The GL coefficients can be obtained in the weak coupling limit within the quasiclassical approximation starting from the nonrelativistic spin-1/2 fermion field theory as \cite{Yasui:2019unp}
\begin{equation}
\begin{alignedat}{2}
& K_0 = \frac{7 \zeta(3) N(0) p_{\rm F}^4}{240 \pi^2 m_{\rm n}^2 T^2},\quad & & \alpha = \frac{N(0) p_{\rm F}^2}{3} \log\frac{T}{T_{\rm c}},\\
& \beta_0 = \frac{7 \zeta(3) N(0) p_{\rm F}^4}{60 \pi^2 T^2},\quad & & \gamma_0 = \frac{31 \zeta(5) N(0) p_{\rm F}^6}{13440 \pi^4 T^4},\\
& \delta_0 = \frac{127 \zeta(7) N(0) p_{\rm F}^8}{387072 \pi^6 T^6},\quad & & \beta_2 = \frac{7 \zeta(3) N(0) p_{\rm F}^2 \gamma_{\rm n}^2}{48 (1 + F_0^a)^2 \pi^2 T^2},\\
& \gamma_2 = \frac{31 \zeta(5) N(0) p_{\rm F}^4 \gamma_{\rm n}^2}{3840 (1 + F_0^a)^2 \pi^4 T^4} & &
\end{alignedat}
\end{equation}
with the temperature $T$, the critical temperature $T_{\rm c}$, the neutron mass $m_{\rm n}$, the neutron gyromagnetic ratio $\gamma_{\rm n}$, the Fermi momentum $p_{\rm F}$, the state-number density $N(0) = m_{\rm n} p_{\rm F} / (2 \pi)^2$ at the Fermi surface, and the Landau parameter $F_0^a$.
\begin{table}
\caption{
\label{table:invariant}
The values of the $\mathrm{U}(1) \times \mathrm{SO}(3)$ invariants $\Vec{S}^2$, $|\Psi_{20}|^2$, and $|\Psi_{30}|^2$
 for ferromagnetic (F), uniaxial nematic (UN), $D_4$ biaxial nematic (BN), $D_2$BN, and cyclic (C) states.
}
\begin{tabular}{cccc}\hline\hline
Phase & $\Vec{S}^2/\rho^2$ & $|\Psi_{20}|^2/\rho^2$ & $|\Psi_{30}|^2/\rho^3$ \\ \hline
F & 4 & 0 & 0 \\
UN & 0 & 1 & 1 \\
$D_4$BN & 0 & 1 & 0 \\
$D_2$BN & 0 & 1 & $(0,1)$ \\
C & 0 & 0 & 2 \\ \hline\hline
\end{tabular}
\end{table}
All uniform states were classified in 
Ref.~\cite{Kobayashi:2021arv}.
The five characteristic symmetric states are ferromagnetic (F), uniaxial nematic (UN), $D_4$ biaxial nematic (BN), $D_2$BN, and cyclic (C) states.
Each uniform state is characterized by $\mathrm{U}(1) \times \mathrm{SO}(3)$ invariants $\Vec{S}^2$, $|\Psi_{20}|^2$, and $|\Psi_{30}|^2$ as summarized in Table \ref{table:invariant}.

For the effective Lagrangian density $f$ in Eq.~\eqref{eq:Lagrangian}, the UN, $D_2$BN, and $D_4$BN states are predicted to be realized 
as the ground states of $^3P_2$ superfluids 
at $|\Vec{B}| = 0$, $0 < |\Vec{B}| < B_{\rm c}$, and $|\Vec{B}| > B_c$,
respectively \cite{Mizushima:2016fbn}, as summarized 
in Table \ref{table:manifold}(a).  
The critical magnetic field $B_{\rm c}$ separating the $D_2$BN and $D_4$BN states depends on the temperature and takes the maximum value $B_{\rm c} = 7.06 \times 10^{-2} \pi (1 + F_0^a) T_{\rm c} / \gamma_{\rm n}$ at $T \simeq 0.854 T_{\rm c}$.
With an estimation for the critical temperature $T_{\rm c} \approx 0.2$ MeV and the Landau parameter $F_0^a \approx 1$, this critical magnetic field can be estimated as $B_{\rm c} \approx 7.36 \times 10^{15}$ G.
At $T \lesssim 0.796 T_{\rm c}$, we obtain $B_{\rm c} = 0$.

\subsection{Single vortex molecule solutions}\label{subsec:single-vortex}

\begin{figure}[htb]
    \centering
    \includegraphics[width=0.99\linewidth]{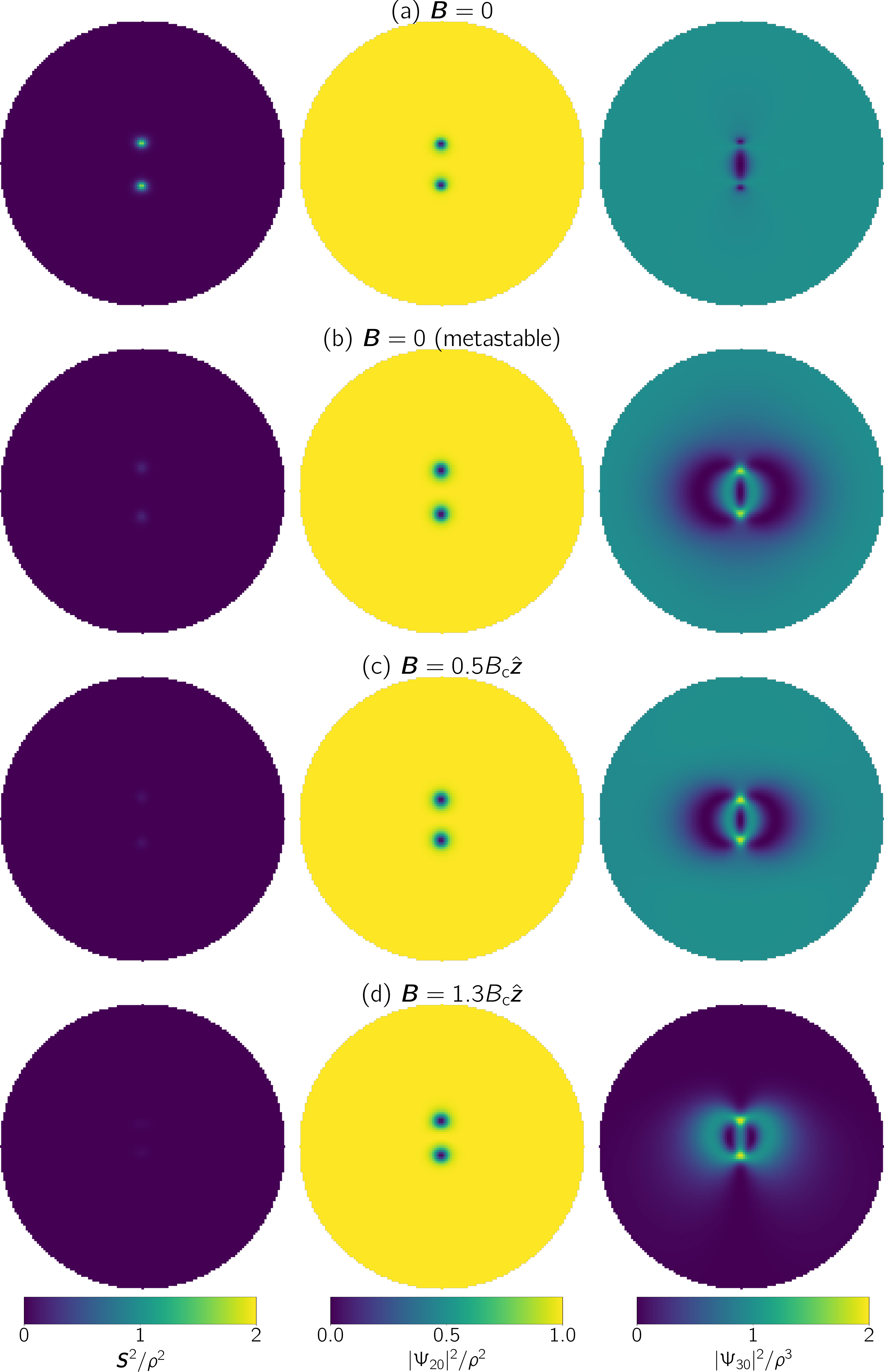}
    \caption{\label{fig:onevortex}
    Singly quantized vortices as molecules of two half-quantized vortices.
    $\Vec{S}^2$ (left panels), $|\Psi_{20}|^2$ (middle panels), and $|\Psi_{30}|^2$ (right panels) for the single-vortex solution at (a) $T=0.854T_{\rm c}$ and $\Vec{B}=0$, (b) $\Vec{B}=0.5B_{\rm c}\hat{\Vec{z}}$, and (c) $\Vec{B}=1.3B_{\rm c}\hat{\Vec{z}}$.
    The radius of the figure shown here is $64p_{\rm F}/(\pi m_{\rm n} T_{\rm c})\approx 9.29$ pm.
    }
\end{figure}

Here, we briefly summarize our previous results 
\cite{Kobayashi:2022moc} 
for singly quantized vortex states 
as molecules of two half-quantized vortices.

First, we consider the ansatz for vortex solutions with vortex cores placed at $r=0$ in the cylindrical coordinates $(r,\theta,z)$ and
the boundary $\psi_{r\to \infty}$ far from vortex cores.
For singly quantized vortices, the order parameters behave as $\psi_{-2\leq m \leq 2}|_{r\to\infty} \propto e^{i\theta}$.
For $\Vec{B}=0$, the uniform ground state is degenerate within the possible UN state
\begin{align}
\begin{aligned}
& \psi_{\pm 2}=\frac{e^{i\phi\mp 2a} \sqrt{3} \sin^2 b}{2\sqrt{2 \rho}}, \quad
\psi_{\pm 1}=\frac{e^{i\phi\mp a} \sin(2b)}{2\sqrt{2 \rho}},\\
& \psi_0=\frac{e^{i\phi}\{1+3\cos(2b)\}}{4 \sqrt{\rho}},
\label{eq:UN-uniform}
\end{aligned}
\end{align}
for $0\leq a,2b,\phi <2\pi$.
Here $a$, $b$, and $\phi$ represent overall spin rotations along the $z$ and $y$ axes, and overall phase shift, respectively.
Under the spatial phase gradient $e^{i\theta}$ for the vortex solution, the current-spin dependent free energy density $f_{202}^{(2)}$ in Eq.~\eqref{eq:Lagrangian} favors $b=0$ giving
\begin{align}
\psi\xrightarrow{r\to\infty}\frac{e^{i\theta}}{\sqrt{\rho}} (0,0,1,0,0)^T.
\label{eq:UN-single-stable}
\end{align}
In the case of $0<|\Vec{B}|<B_{\rm c}$, we obtain
\begin{align}
\psi\xrightarrow{r\to\infty}\frac{e^{i\theta}}{\sqrt{\rho}} \left(\frac{e^{-2 i a}\sin g}{\sqrt{2}},0,\cos g,0,\frac{e^{2 i a}\sin g}{\sqrt{2}}\right)^T,
\label{eq:D2BN-single}
\end{align}
where $g$ depends on $|\Vec{B}|$ and satisfies $\pi/3 < g < \pi/2$, making $\psi|_{r\to\infty}$ to be the $D_2$BN state.
$a$ also represents the overall spin rotation along the $z$ axis and takes arbitrary (fixed) value without (with) the current-spin dependent free energy $f_{202}^{(1)}$.
In the limit of $|\Vec{B}|\searrow 0$, $g$ becomes $g\to \pi/3$ giving
\begin{align}
\psi\xrightarrow{r\to\infty}\frac{e^{i\theta}}{\sqrt{\rho}} \left(\frac{e^{-2 i a}\sqrt{3}}{2\sqrt{2}},0,\frac{1}{2},0,\frac{e^{2 i a}\sqrt{3}}{2\sqrt{2}}\right)^T.
\label{eq:UN-single-metastable}
\end{align}
This solution belongs to the UN state in Eq.~\eqref{eq:UN-uniform} with $b=\pi/2$, but is different from that for $\Vec{B}=0$ shown in Eq.~\eqref{eq:UN-single-stable},
which leads the discontinuity between $\Vec{B}=0$ and $|\Vec{B}|\searrow 0$ and the metastable solution at $\Vec{B}=0$.
For the case of $|\Vec{B}|\geq B_{\rm c}$, $g$ becomes $g=\pi/2$, giving
\begin{align}
\psi\xrightarrow{r\to\infty}\frac{e^{i\theta}}{\sqrt{\rho}} \left(\frac{e^{-2 i a}}{\sqrt{2}},0,0,0,\frac{e^{2 i a}}{\sqrt{2}}\right)^T,
\label{eq:D4BN-single}
\end{align}
which belongs to the $D_4$BN state.

Next, we show our numerical results for singly quantized vortex solutions.
They are obtained by minimizing the free-energy density $f$ under the boundary conditions with the cylindrical coordinates $(r,\theta,z)$:
\begin{align}
\left.\psi_m(\theta+\pi)\right|_{r=L}=-\left.\psi_m(\theta)\right|_{r=L},
\label{eq:boundary_condition_single_vortex}
\end{align}
at the boundary $r=L$, which induces a singly quantized vortex solution.
The minimization of the free energy density $f$ can be done by finding the stationary solution of the GL equation
\begin{align}
\frac{\delta f}{\delta \psi_m^\ast}=0.
\label{eq:GL_single_zero}
\end{align}
The solution of Eq.~\eqref{eq:GL_single_zero} can be obtained by the Nesterov's method with introducing the relaxation time $t$ and the dependence of the order parameter $\psi_m$.
The time dependences of $\psi_m$ is given by
\begin{align}
\begin{aligned}
\ddot{\psi}_m = - \frac{\delta f}{\delta \psi_m^\ast} - \frac{3}{t} \dot{\psi}_m, \quad
\dot{\psi}_m(t=0) = 0.
\label{eq:Nesterov}
\end{aligned}
\end{align}
After the long time evolution of Eq.~\eqref{eq:Nesterov}, we obtain the solution of Eq.~\eqref{eq:GL_single_zero}.
We note that Eq.~\eqref{eq:Nesterov} is just one of methods to effectively obtain solutions to Eq.~\eqref{eq:GL_single_zero}.

Figure~\ref{fig:onevortex} shows numerical solutions 
\cite{Kobayashi:2022moc} at the temperature $T=0.854 T_{\rm c}$ and the magnetic field $\Vec{B}=0$, $0.5B_{\rm c}\hat{\Vec{z}}$, and $1.3B_{\rm c}\hat{\Vec{z}}$.
In any cases, a singly quantized vortex splits into two half-quantized vortices with holes of $|\Psi_{20}|^2$ forming a vortex molecule.
The fact that isolated half-quantized vortices can topologically exist only in the $D_4$BN state implies that the $D_4$BN order should appear around the vortex core for even for 
$|\Vec{B}|<B_{\rm c}$ 
in which the ground states are either UN or $D_2$BN states 
[Figs.~\ref{fig:onevortex}(a)--(c)].
In fact, we can confirm that the $D_4$BN order characterized by $\Vec{S}^2 = 0$, $|\Psi_{20}|^2/\rho^2 = 1$, and $|\Psi_{30}|^2/\rho^3 = 0$ appears as one (three) soliton(s) bridging two vortex cores for Fig.~\ref{fig:onevortex}(a) at $\Vec{B}=0$ [Fig.~\ref{fig:onevortex} (b) at $\Vec{B}=0$ and Fig.~\ref{fig:onevortex} (c) at $\Vec{B}=0.5B_{\rm c}\hat{\Vec{z}}$], as can be seen in the plot of $|\Psi_{30}|^2$ locally inducing the $D_4$BN order.
At $\Vec{B}=1.3 B_{\rm c}\hat{\Vec{z}}$ shown in Fig.~\ref{fig:onevortex}(d), where $D_4$BN state becomes the ground state, two half-quantized vortices also form a vortex molecule bridged by three $D_2$BN solitons characterized by $0 < |\Psi_{30}|^2/\rho^3<1$.
The cores of the half-quantized vortices are filled by F, C, C, and C orders for Figs.~\ref{fig:onevortex}(a), \ref{fig:onevortex}(b), \ref{fig:onevortex}(c), and \ref{fig:onevortex}(d), respectively.
At the numerical boundary $r=L$, the order parameters for Figs.~\ref{fig:onevortex}(a), \ref{fig:onevortex}(b), \ref{fig:onevortex}(c), and \ref{fig:onevortex}(d) satisfy Eqs.~\eqref{eq:UN-single-stable}, \eqref{eq:UN-single-metastable}, \eqref{eq:D2BN-single}, and \eqref{eq:D4BN-single}, respectively.
In particular, the two solutions at $\Vec{B}=0$ shown in Figs.~\ref{fig:onevortex}(a) and (b) correspond to the most stable and metastable ones, respectively, 
and the latter was missing in the previous study 
\cite{Kobayashi:2022moc}.

\section{Multi vortex state for $^3P_2$ Neutron superfluids}\label{sec:multivortex}

In this section, we show our main results for multi vortex state for $^3P_2$ superfluids.
Because two vortices with the same circulation have long-range repulsion, they cannot be stabilized by the boundary condition such as 
Eq.~\eqref{eq:boundary_condition_single_vortex}. 
We thus introduce an external rotation term by rewriting the free-energy density $f$ as $f - \mathrm{Re}[\Vec{\Omega} \cdot \Vec{L}]$ \cite{salomaaPRB85},
where $\Vec{\Omega}=(0,0,\Omega)$ is the rotation vector parallel to the $z$ axis, and $\Vec{L} = m_{\rm n}^2 p_{\rm F} \psi^\dagger(-i\Vec{r}\times \nabla)\psi$ is the angular momentum.
This is nothing but a physical situation realized in 
rotating neutron stars.
As well as the case of the single-vortex solutions, we obtain the stationary solution of the GL equation by the Nesterov's method in 
Eq.~\eqref{eq:Nesterov} under the boundary condition
\begin{align}
    \left. \left(\nabla \psi\right)_\parallel\right|_{r=L}=0,
\end{align}
where $(\ )_\parallel$ means the component parallel to $\hat{\Vec{r}}$ in the cylindrical coordinate, and $L$ is the numerical boundary set to be $L=128p_{\rm F}/(\pi m_{\rm n} T_{\rm c}) \approx 18.6$ pm.

\subsection{Two-vortex molecule states}

\subsubsection{The stable solution in the case of zero magnetic field $\Vec{B}=0$}

\begin{figure*}[htb]
    \centering
    \includegraphics[width=0.77\linewidth]{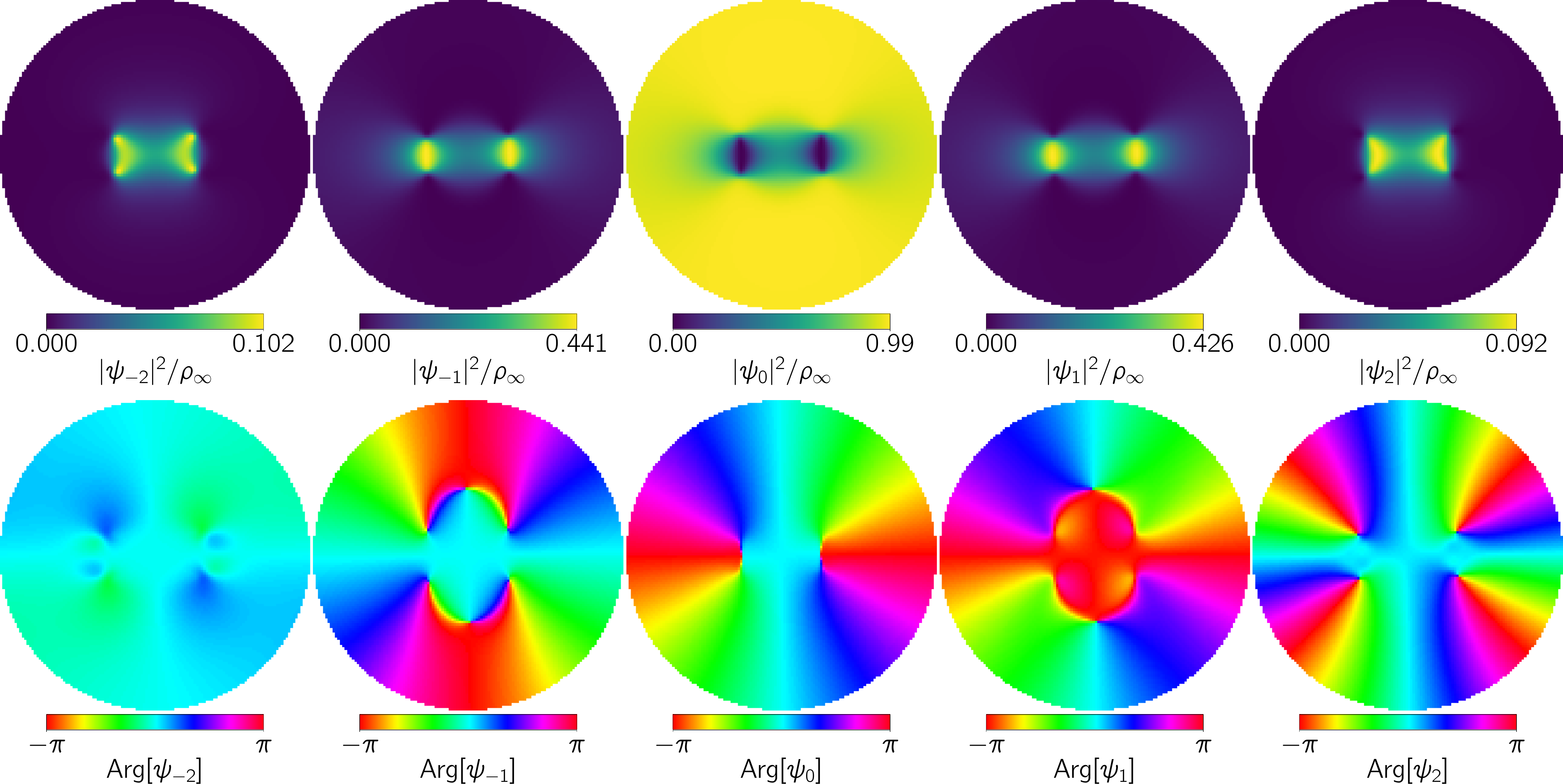}
    \includegraphics[width=0.77\linewidth]{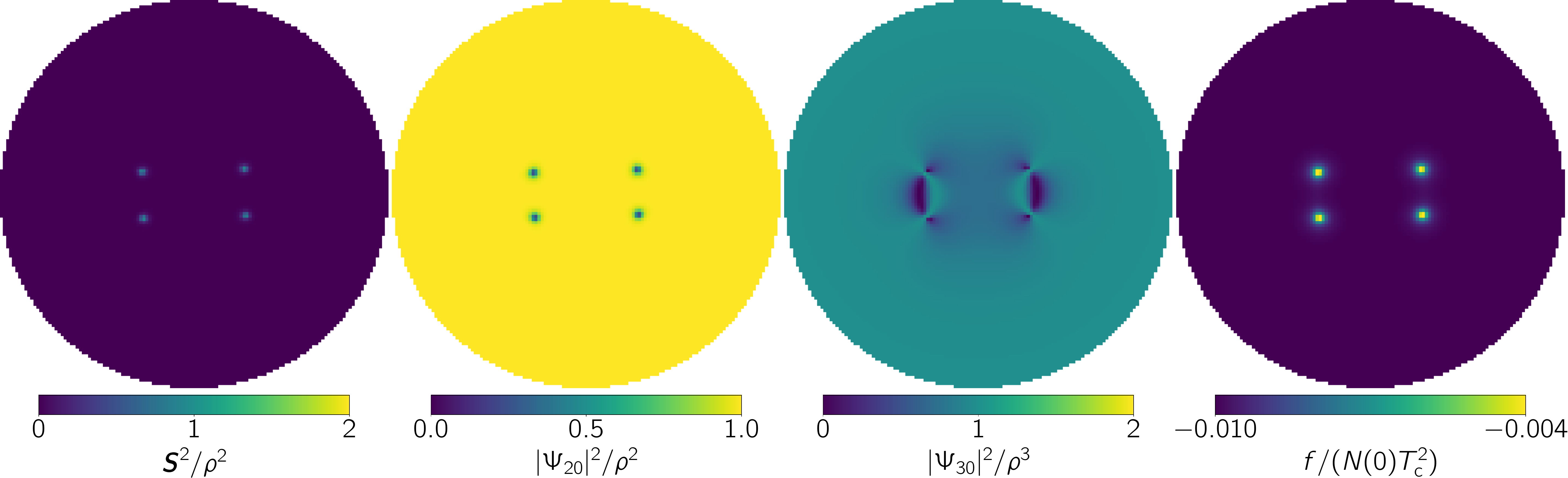}
    \caption{
        \label{fig:vortex-85-000-02-08}
        The most stable two-vortex molecule state in the UN phase at $\Vec{B}=0$ and $\Omega^2=0.8\Omega_0^2$ with $\Omega_0=160 \pi^2 m_n T_{\rm c}^2/(256^2 p_{\rm F}^2)\approx 7.15\times 10^{15}$ rad\ fs$^{-1}$.
        The squared modulus $|\psi_m|^2$ (top row), argument $\mathrm{Arg}[\psi_m]$ (middle row) of the order parameter, the $\mathrm{U}(1) \times \mathrm{SO}(3)$ invariants $\Vec{S}^2$, $|\Psi_{20}|^2$, and $|\Psi_{30}|^2$, and the free-energy density $f$ (bottom row) are shown. 
        The radius of the figure shown here is $64 p_{\rm F} / (\pi m_{\rm n} T_{\rm c}) \approx 9.29$ pm.
    }
\end{figure*}
For the $\Vec{B}=0$ case in which the UN phase is the ground state, we expect the F-core vortex molecules connected by a single $D_4$BN soliton as shown in Fig.~\ref{fig:onevortex}(a).
Figure~\ref{fig:vortex-85-000-02-08} shows the two-vortex molecule state at $\Vec{B}=0$ and $\Omega^2=0.8\Omega_0^2$, where $\Omega_0$ is defined as $\Omega_0=160 \pi^2 m_n T_{\rm c}^2/(256^2 p_{\rm F}^2)\approx 7.30$ rad\ fs$^{-1}$.
Compared to the singly quantized vortex state, the two $D_4$BN solitons between half-quantized vortex cores repel each other to bend,  
as can be clearly seen in $|\Psi_{30}|^2$.
At the boundary $r=L$, the order parameter approximately satisfies
\begin{align}
\psi |_{r=L}=\frac{e^{2i\theta}}{\sqrt{\rho}} (0,0,1,0,0)^T,
\label{eq:UN-double-stable}
\end{align}
which is just the double winding of the singly quantized vortex solution Eq.~\eqref{eq:UN-single-stable} at $\Vec{B}=0$.

\begin{figure*}[htb]
    \centering
    \includegraphics[width=0.77\linewidth]{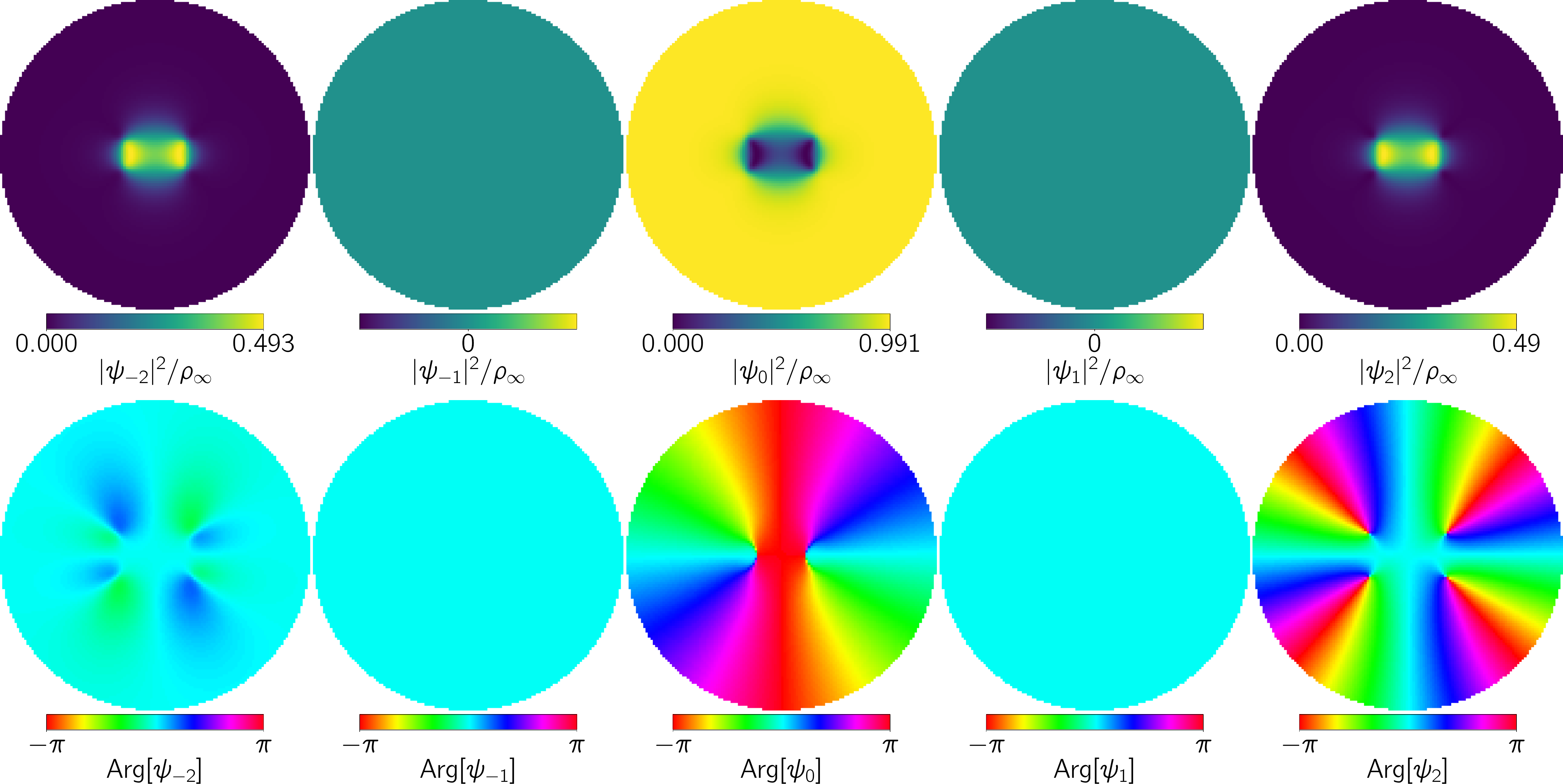}
    \includegraphics[width=0.77\linewidth]{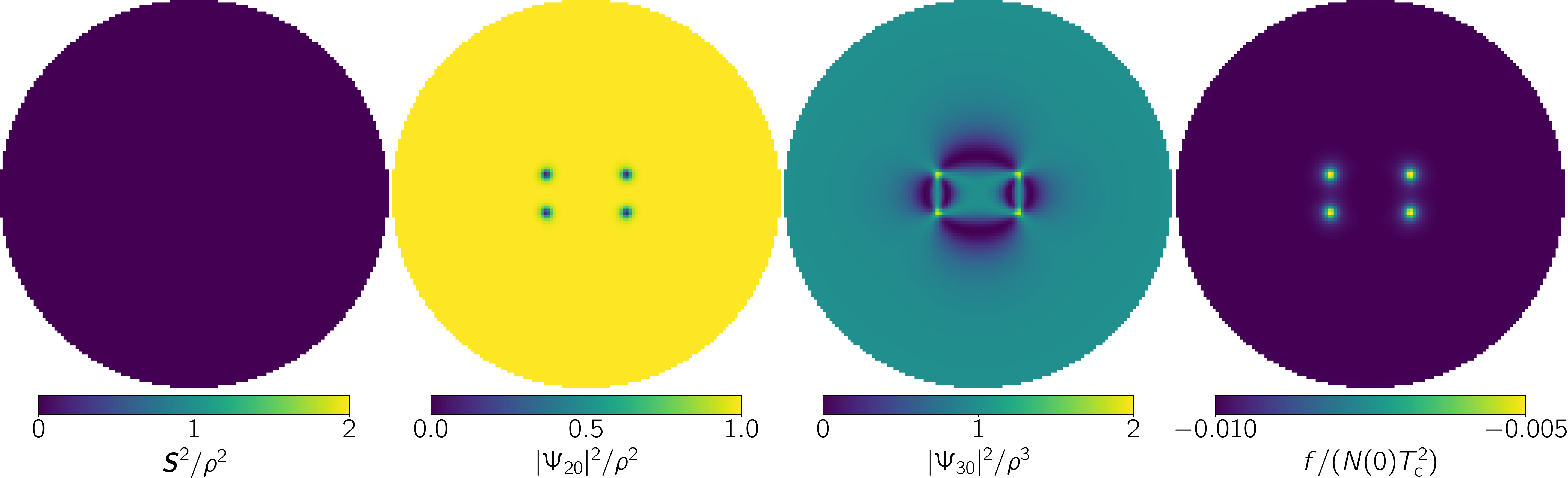}
    \caption{
        \label{fig:vortex-85-000-02-10}
        The most stable two-vortex molecule state in the UN phase at $\Vec{B}=0$ and $\Omega^2=1.0\Omega_0^2$.
        The squared modulus $|\psi_m|^2$ (top row), argument $\mathrm{Arg}[\psi_m]$ (middle row) of the order parameter, the $\mathrm{U}(1) \times \mathrm{SO}(3)$ invariants $\Vec{S}^2$, $|\Psi_{20}|^2$, and $|\Psi_{30}|^2$, and the free-energy density $f$ (bottom row) are shown. 
        The radius of the figure shown here is $64 p_{\rm F} / (\pi m_{\rm n} T_{\rm c}) \approx 9.29$ pm.
        The two vortex molecules share two solitons.
    }
\end{figure*}
With increasing the rotation, the ``dimerization'' of two vortex molecules occurs while keeping the boundary state unchanged as shown in Eq.~\eqref{eq:UN-double-stable}.
Figure~\ref{fig:vortex-85-000-02-10} shows the two-vortex molecule 
state 
consisting of four half-quantized vortices 
at $\Omega^2=1.0\Omega_0^2$.
The details of the dimerization is as follows.
First, the F-core vortex molecules having one $D_4$BN soliton changes to C-core vortex molecules having three $D_4$BN solitons 
as in the case of the vortex molecule at $0 < |\Vec{B}| < B_{\rm c}$. 
Then, one of three $D_4$BN solitons reconnect to one of the other molecules to be shared by the two vortex molecules to form 
a ``covalent bond''.
In Fig.~\ref{fig:vortex-85-000-02-10}, 
the upper and lower $D_4$BN solitons in $|\Psi_{30}|^2$ are covalent bonds between left and right vortex molecules.
The transition from the two isolated F-core vortex molecules to a dimerized C-core vortex molecule occurs at $\Omega^2 \simeq \Omega_0^2$.

\subsubsection{The metastable solution in the case of zero magnetic field $\Vec{B}=0$}

\begin{figure*}[htb]
    \centering
    \includegraphics[width=0.77\linewidth]{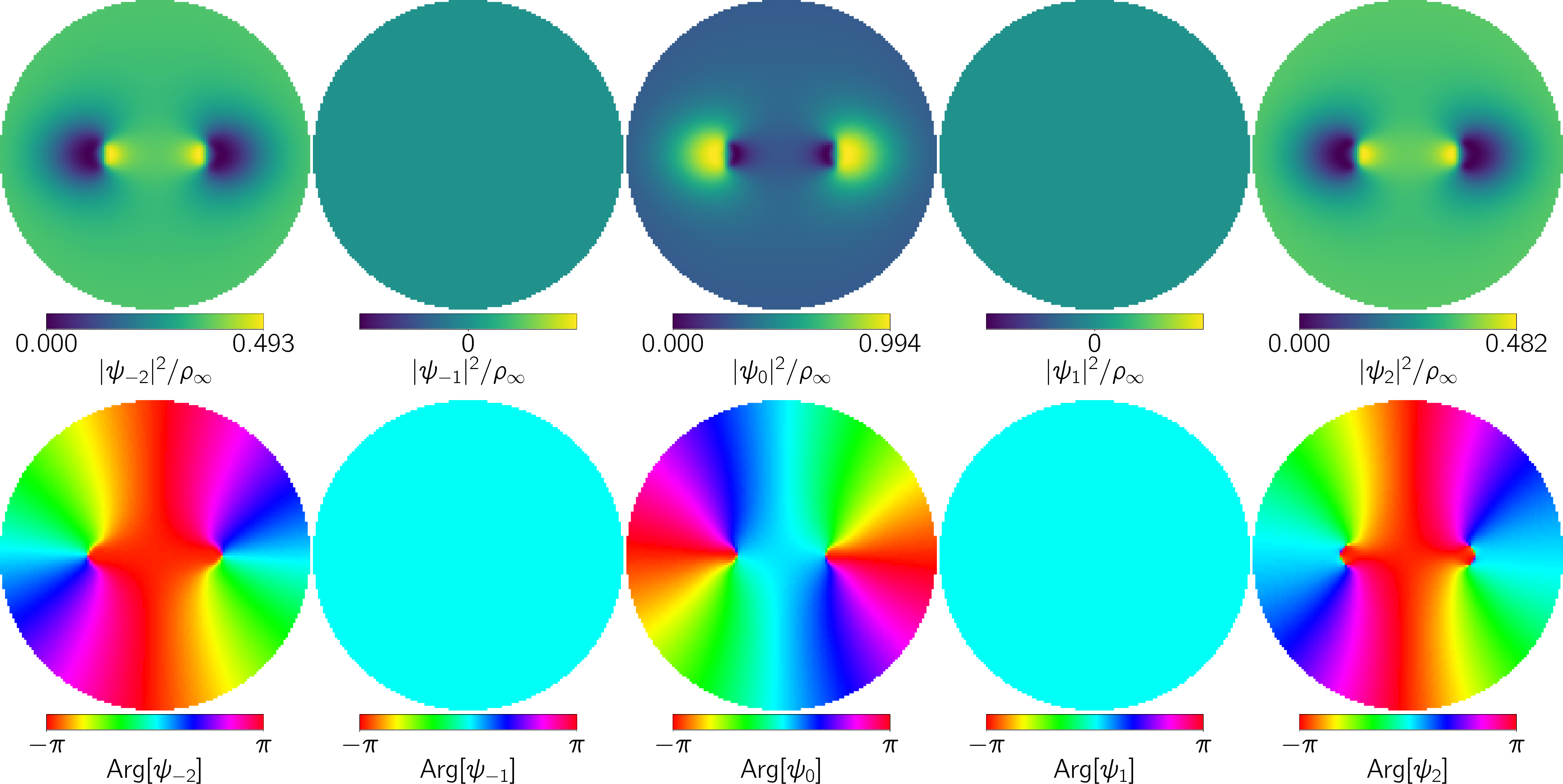}
    \includegraphics[width=0.77\linewidth]{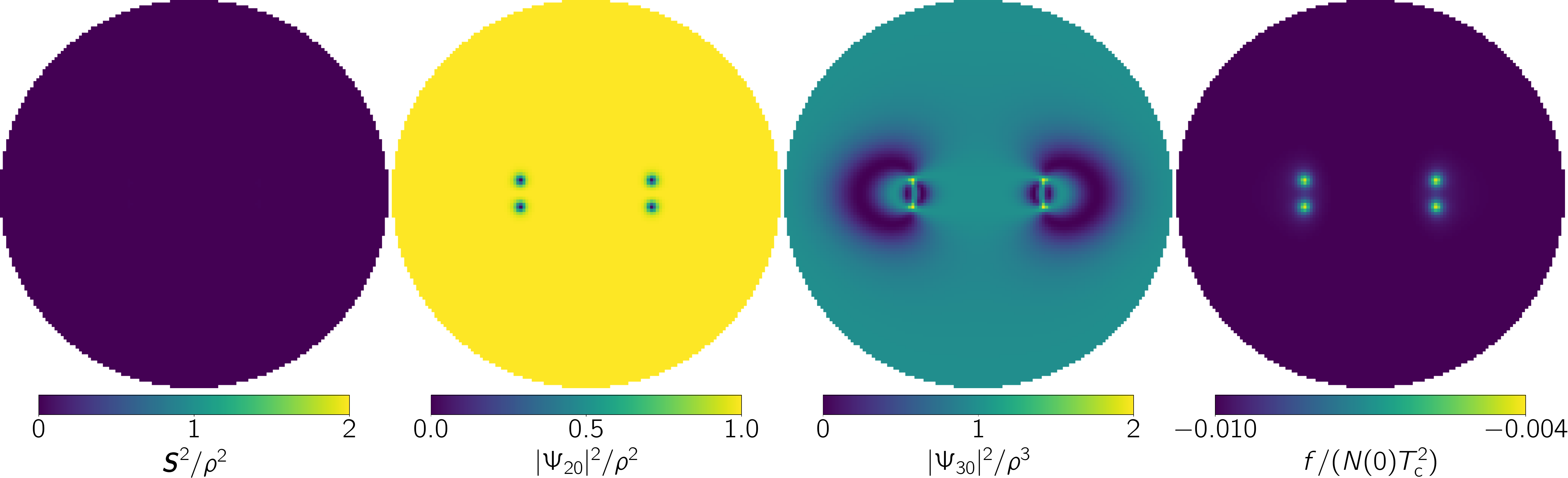}
    \caption{
        \label{fig:vortex-85-000-02-08-ms}
        The metastable two-vortex molecule state in the UN phase at $\Vec{B}=0$ and $\Omega^2=0.8\Omega_0^2$.
        The squared modulus $|\psi_m|^2$ (top row), argument $\mathrm{Arg}[\psi_m]$ (middle row) of the order parameter, the $\mathrm{U}(1) \times \mathrm{SO}(3)$ invariants $\Vec{S}^2$, $|\Psi_{20}|^2$, and $|\Psi_{30}|^2$, and the free-energy density $f$ (bottom row) are shown. 
        The radius of the figure shown here is $64 p_{\rm F} / (\pi m_{\rm n} T_{\rm c}) \approx 9.29$ pm.
    }
\end{figure*}
As well as singly quantized vortex states, there are metastable solutions having the other boundary
\begin{align}
\psi |_{r=L}=\frac{e^{2i\theta}}{\sqrt{\rho}} \left(\frac{e^{-2 i a}\sqrt{3}}{2\sqrt{2}},0,\frac{1}{2},0,\frac{e^{2 i a}\sqrt{3}}{2\sqrt{2}}\right)^T,
\label{eq:UN-double-metastable}
\end{align}
which is double winding of Eq.~\eqref{eq:UN-single-metastable}.
In this case, we expect the C-core vortex molecule connected by three $D_4$BN solitons as shown in Fig.~\ref{fig:onevortex}(b).
Figure~\ref{fig:vortex-85-000-02-08-ms} shows the metastable two-vortex molecule state at $\Vec{B}=0$ and $\Omega^2=0.8\Omega_0^2$.
There are two isolated vortex molecules in which the three $D_4$BN solitons connect half-quantized vortices.
Compared to the single vortex state shown in Fig.~\ref{fig:onevortex}(b), the $D_4$BN soliton in the side of the other vortex molecule is shorten.
On the other hand, the $D_4$BN soliton in the opposite side is enlarged and bent.
As a result, a symmetric structure of one vortex molecule as shown in Fig.~\ref{fig:onevortex}(b) is strongly distorted.

\begin{figure*}[htb]
    \centering
    \includegraphics[width=0.77\linewidth]{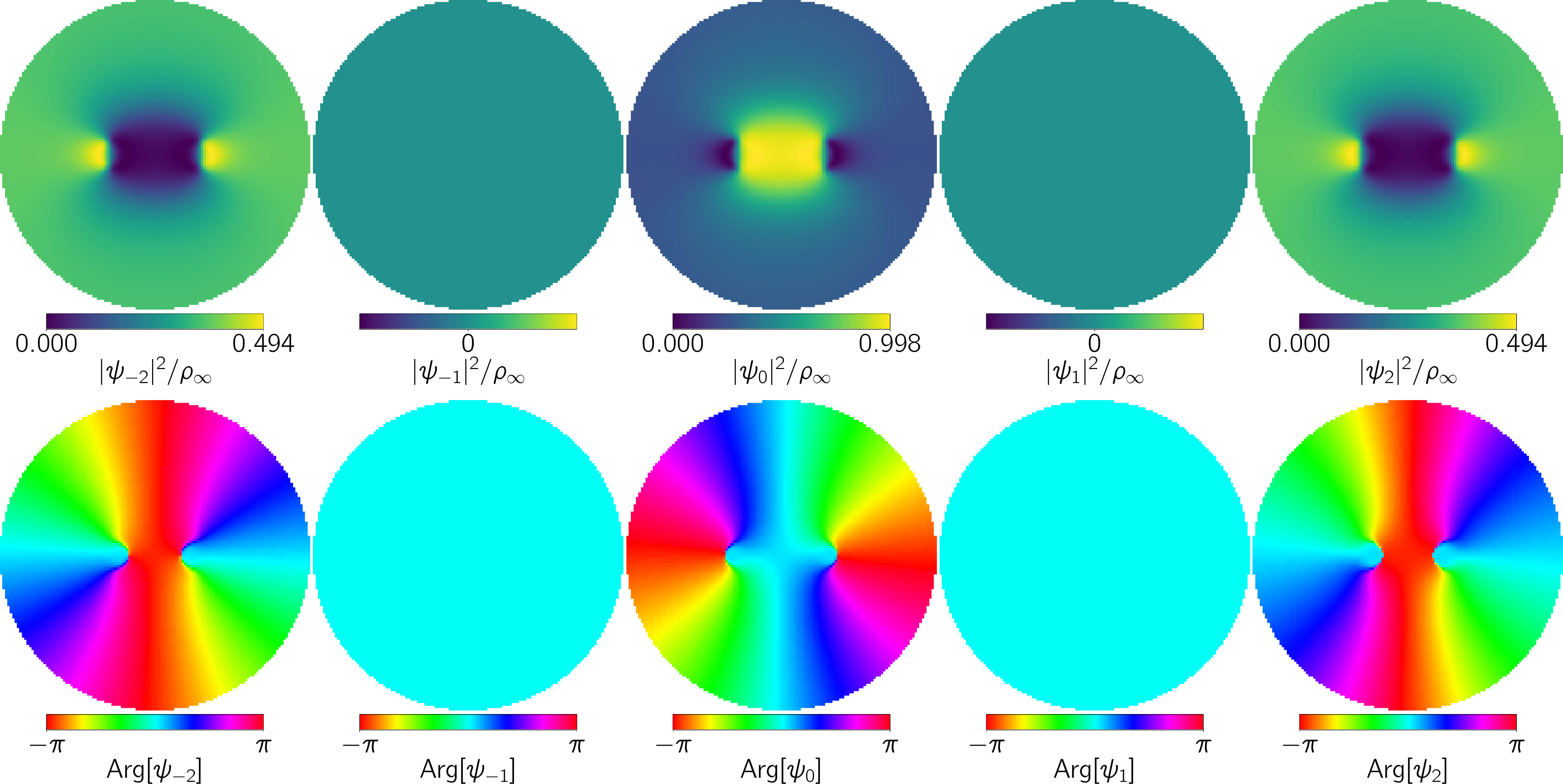}
    \includegraphics[width=0.77\linewidth]{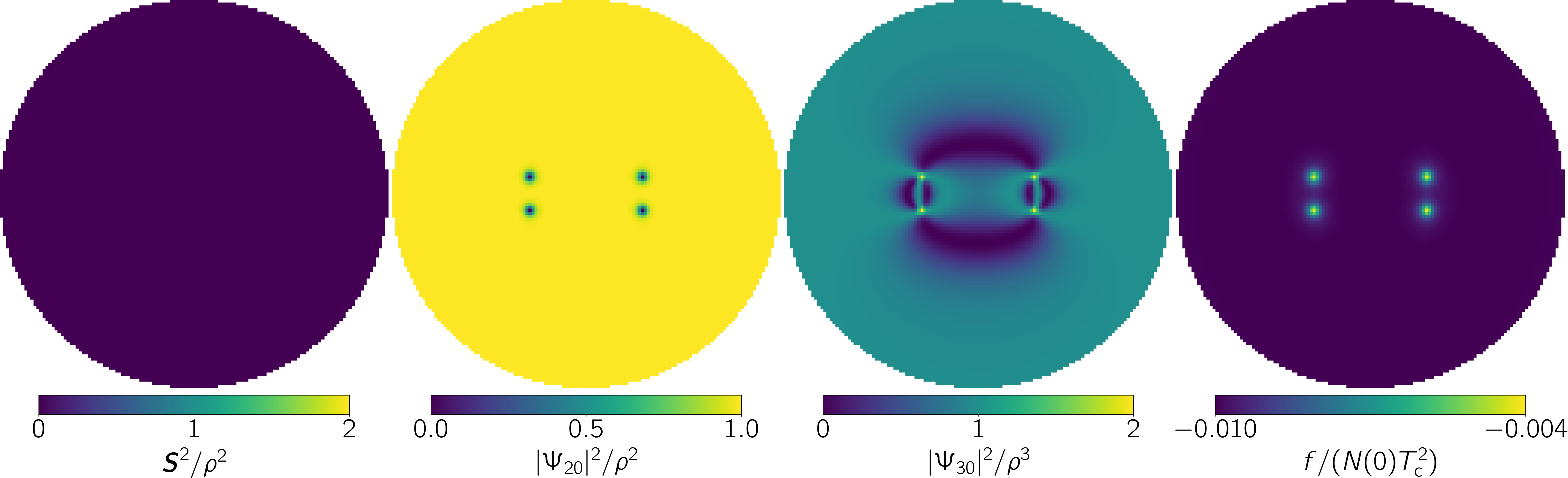}
    \caption{
        \label{fig:vortex-85-000-02-10-ms}
        The metastable two-vortex molecule state in the UN phase at $\Vec{B}=0$ and $\Omega^2=1.0\Omega_0^2$.
        The squared modulus $|\psi_m|^2$ (top row), argument $\mathrm{Arg}[\psi_m]$ (middle row) of the order parameter, the $\mathrm{U}(1) \times \mathrm{SO}(3)$ invariants $\Vec{S}^2$, $|\Psi_{20}|^2$, and $|\Psi_{30}|^2$, and the free-energy density $f$ (bottom row) are shown. 
        The radius of the figure shown here is $64 p_{\rm F} / (\pi m_{\rm n} T_{\rm c}) \approx 9.29$ pm.
    }
\end{figure*}
As well as the stable case, increasing $\Omega$ causes the dimerization of vortex molecules under the same boundary condition Eq.~\eqref{eq:UN-double-metastable}.
The shape of the dimerized vortex molecule is similar to that shown in Fig.~\ref{fig:vortex-85-000-02-10}, i.e., there are two $D_4$BN covalent bonds shared by the two vortex molecules.

\subsubsection{The case of 
a small magnetic field $0 < |\Vec{B}| < B_{\rm c}$}

\begin{figure*}[htb]
    \centering
    \includegraphics[width=0.77\linewidth]{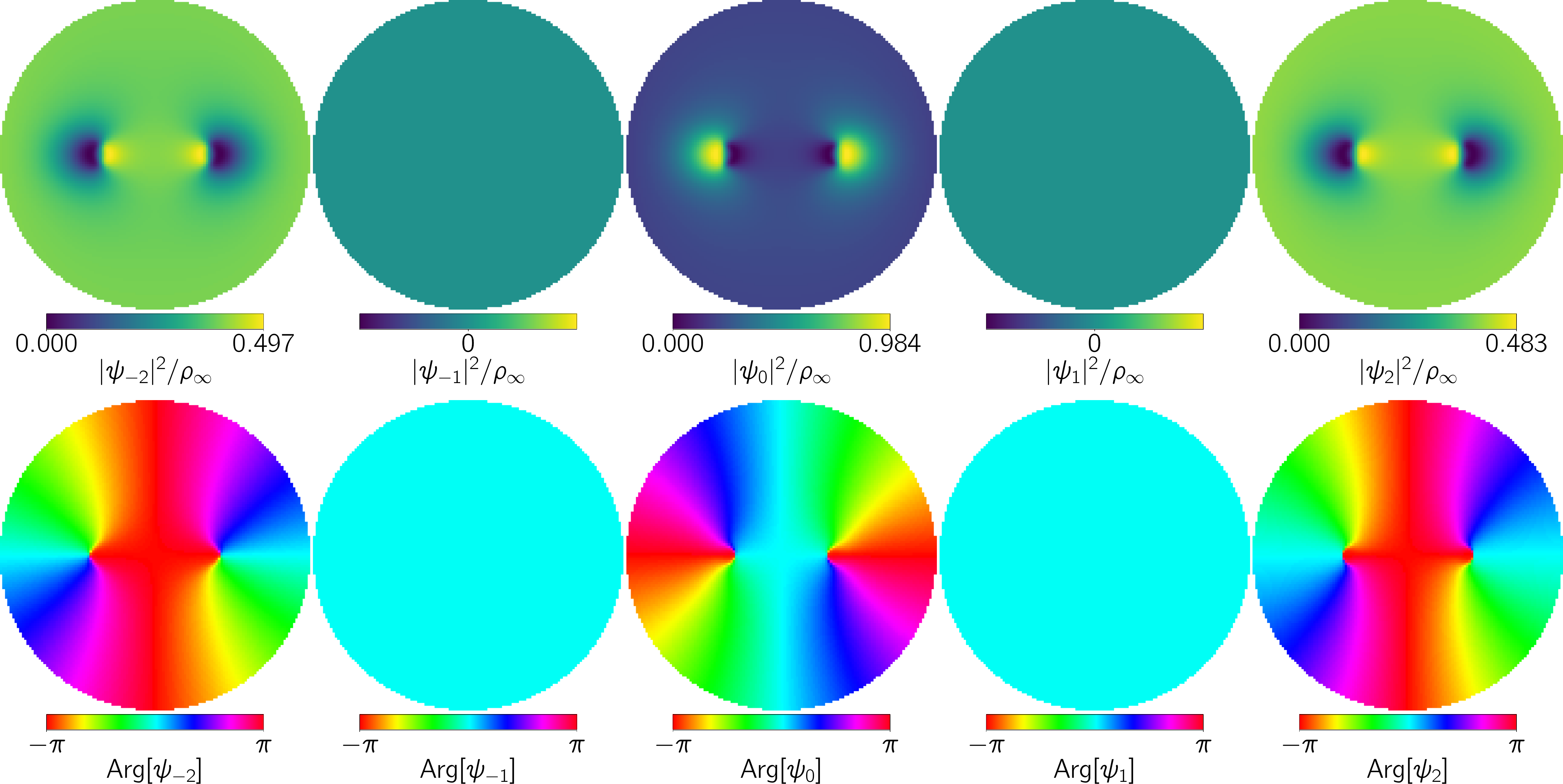}
    \includegraphics[width=0.77\linewidth]{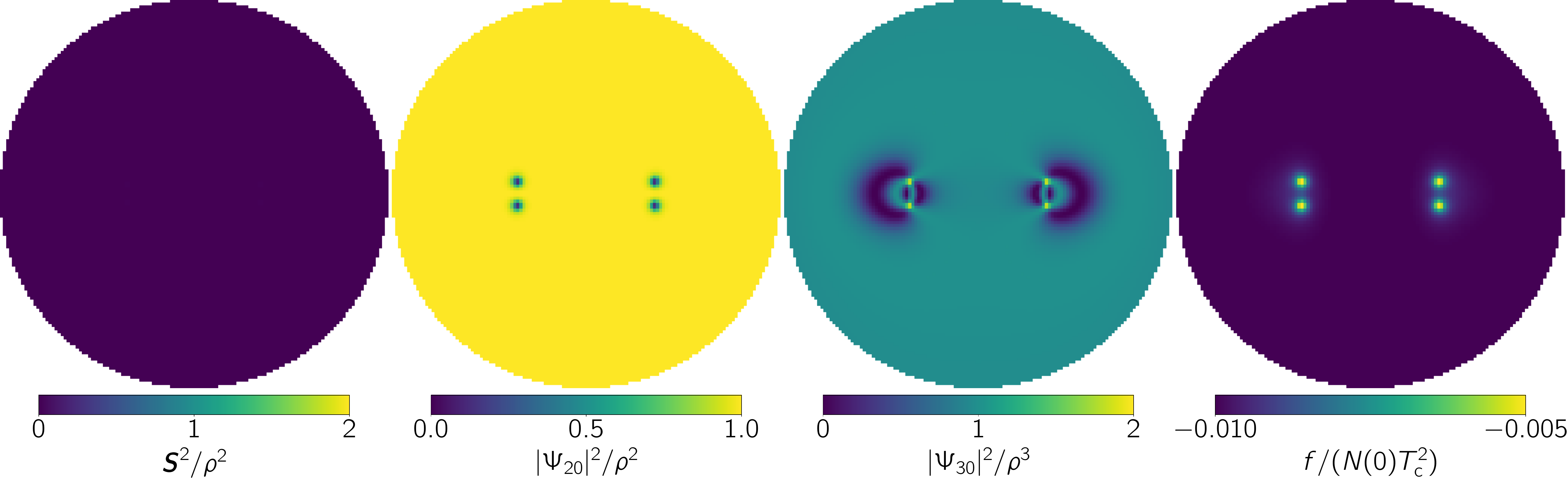}
    \caption{
        \label{fig:vortex-85-050-02-08}
        The two-vortex molecule state in the $D_2$BN phase at $\Vec{B}=0.5B_{\rm c}\hat{\Vec{z}}$ and $\Omega^2=0.8\Omega_0^2$.
        The squared modulus $|\psi_m|^2$ (top row), argument $\mathrm{Arg}[\psi_m]$ (middle row) of the order parameter, the $\mathrm{U}(1) \times \mathrm{SO}(3)$ invariants $\Vec{S}^2$, $|\Psi_{20}|^2$, and $|\Psi_{30}|^2$, and the free-energy density $f$ (bottom row) are shown. 
        The radius of the figure shown here is $64 p_{\rm F} / (\pi m_{\rm n} T_{\rm c}) \approx 9.29$ pm.
    }
\end{figure*}

\begin{figure*}[htb]
    \centering
    \includegraphics[width=0.77\linewidth]{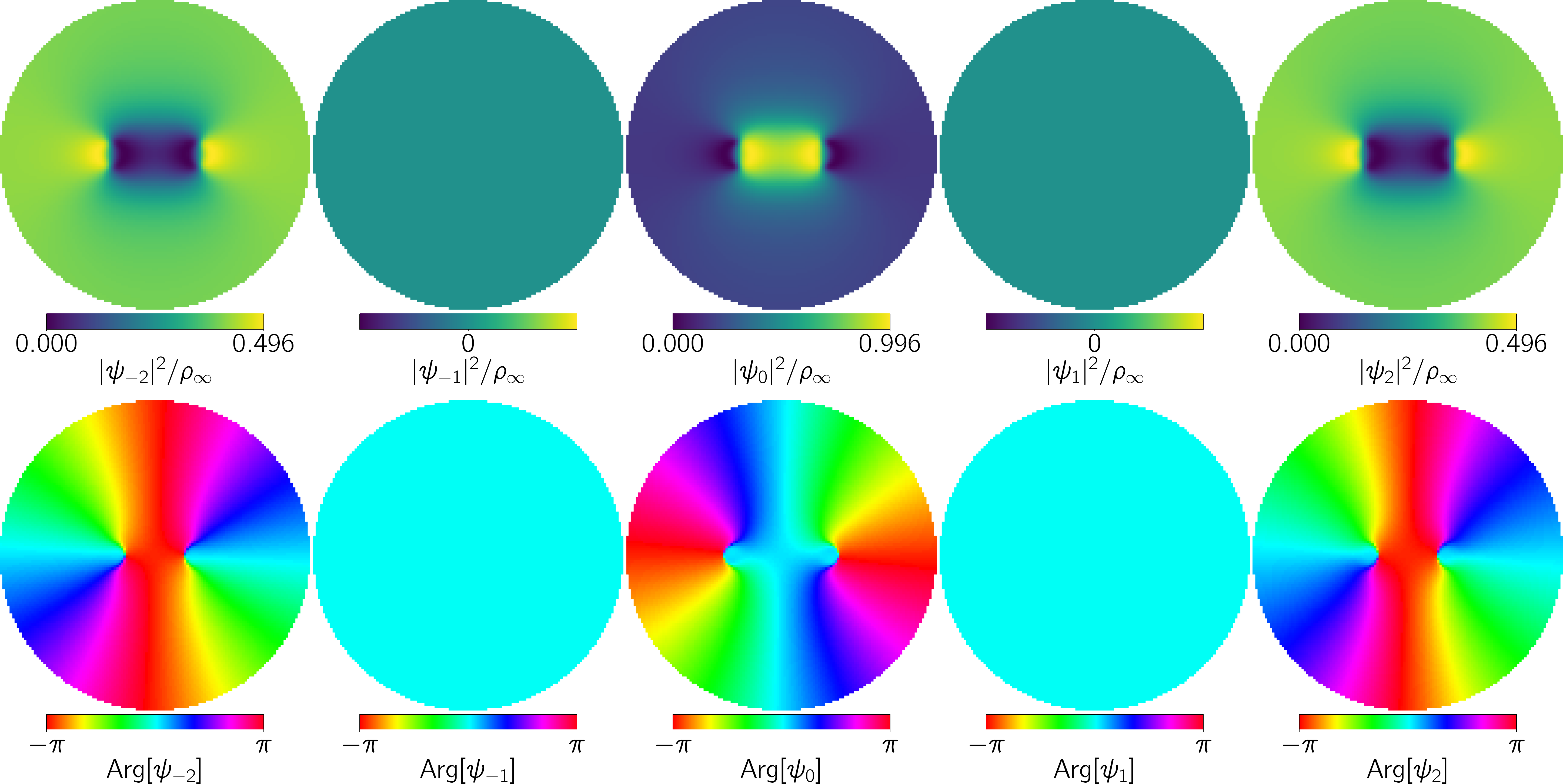}
    \includegraphics[width=0.77\linewidth]{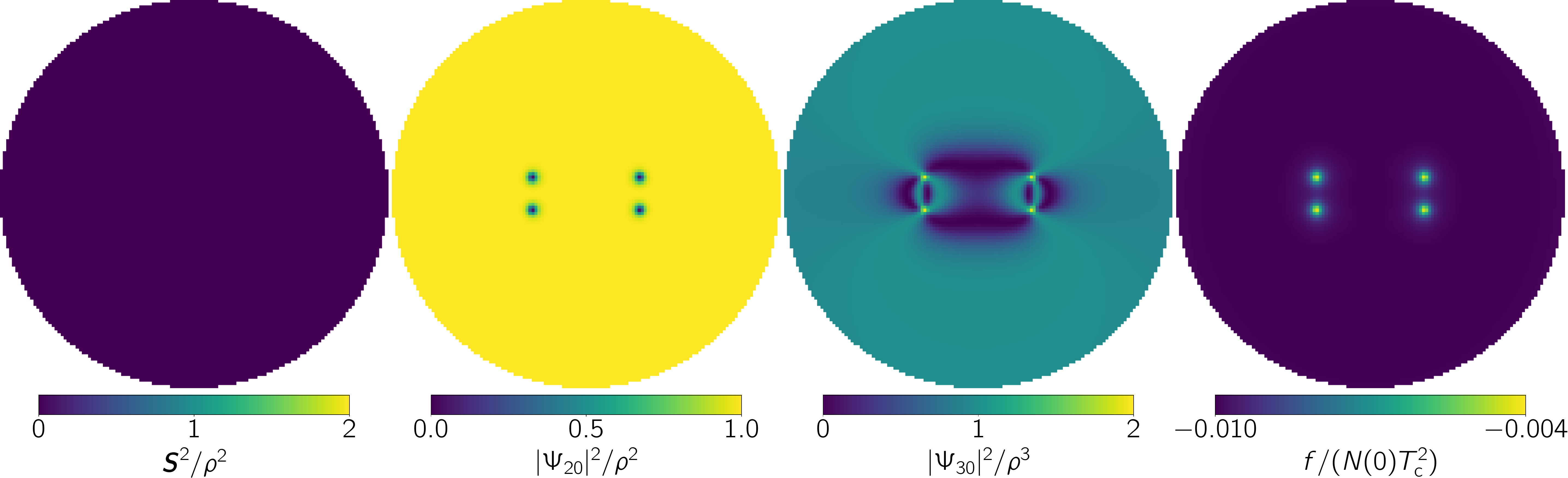}
    \caption{
        \label{fig:vortex-85-050-02-10}
        The two-vortex molecule state in the $D_2$BN phase at $\Vec{B}=0.5B_{\rm c}\hat{\Vec{z}}$ and $\Omega^2=1.0\Omega_0^2$.
        The squared modulus $|\psi_m|^2$ (top row), argument $\mathrm{Arg}[\psi_m]$ (middle row) of the order parameter, the $\mathrm{U}(1) \times \mathrm{SO}(3)$ invariants $\Vec{S}^2$, $|\Psi_{20}|^2$, and $|\Psi_{30}|^2$, and the free-energy density $f$ (bottom row) are shown. 
        The radius of the figure shown here is $64 p_{\rm F} / (\pi m_{\rm n} T_{\rm c}) \approx 9.29$ pm.
    }
\end{figure*}
For the $0<|\Vec{B}|<B_{\rm c}$ case in which the $D_2$BN phase is the ground state, we expect the C-core vortex molecule connected by three $D_4$BN solitons as shown in Fig.~\ref{fig:onevortex}(c).
Figure~\ref{fig:vortex-85-050-02-08} (Fig.~\ref{fig:vortex-85-050-02-10}) shows two-vortex molecule state at $\Vec{B}=0.5B_{\rm c}\hat{\Vec{z}}$ and $\Omega^2=0.8\Omega_0^2$ ($\Omega^2=1.0\Omega_0^2$).
Two characteristic structures for two isolated vortex molecules shown in Fig.~\ref{fig:vortex-85-050-02-08} and the dimerized vortex molecule shown in Fig.~\ref{fig:vortex-85-050-02-10} are almost the same as those for the metastable solutions at $\Vec{B}=0$ for Figs.~\ref{fig:vortex-85-000-02-08-ms} and \ref{fig:vortex-85-000-02-10-ms}, while the order parameter at the boundary satisfy
\begin{align}
\psi |_{r=L}=\frac{e^{2i\theta}}{\sqrt{\rho}} \left(\frac{e^{-2 i a}\sin g}{\sqrt{2}},0,\cos g,0,\frac{e^{2 i a}\sin g}{\sqrt{2}}\right)^T,
\label{eq:D2BN-double}
\end{align}
which is double winding of Eq.~\eqref{eq:D2BN-single}.
As well as the singly quantized vortex solution, the two-vortex molecule state in the $D_2$BN phase at $0<|\Vec{B}|<B_{\rm c}$ continuously changes to the metastable state in the UN phase at $\Vec{B}=0$ [$g\to\pi/3$ limit in Eq.~\eqref{eq:D2BN-double} at the boundary].


\subsubsection{The case of large magnetic field 
$|\Vec{B}|\geq B_{\rm c}$}

\begin{figure*}[htb]
    \centering
    \includegraphics[width=0.77\linewidth]{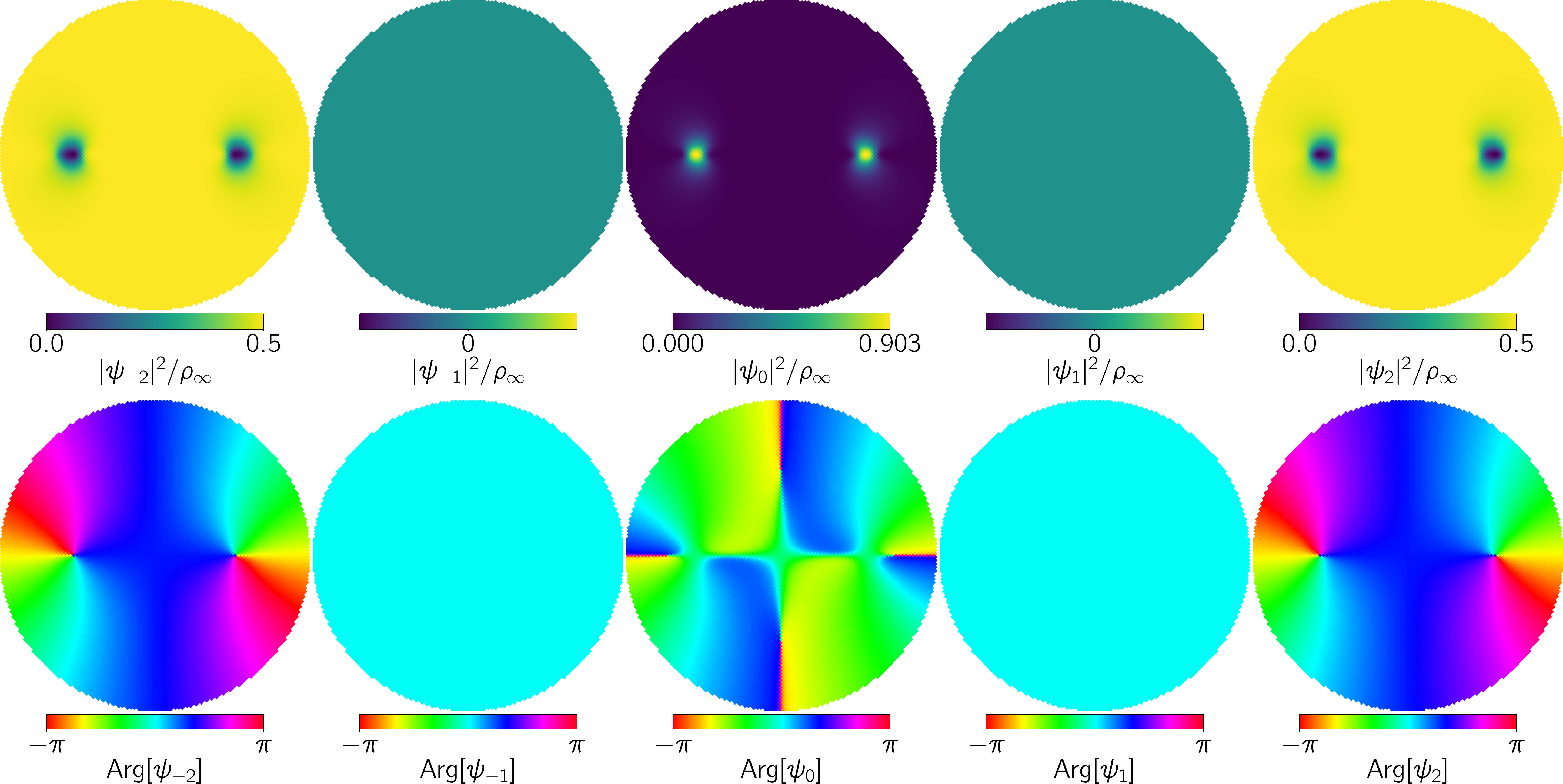}
    \includegraphics[width=0.77\linewidth]{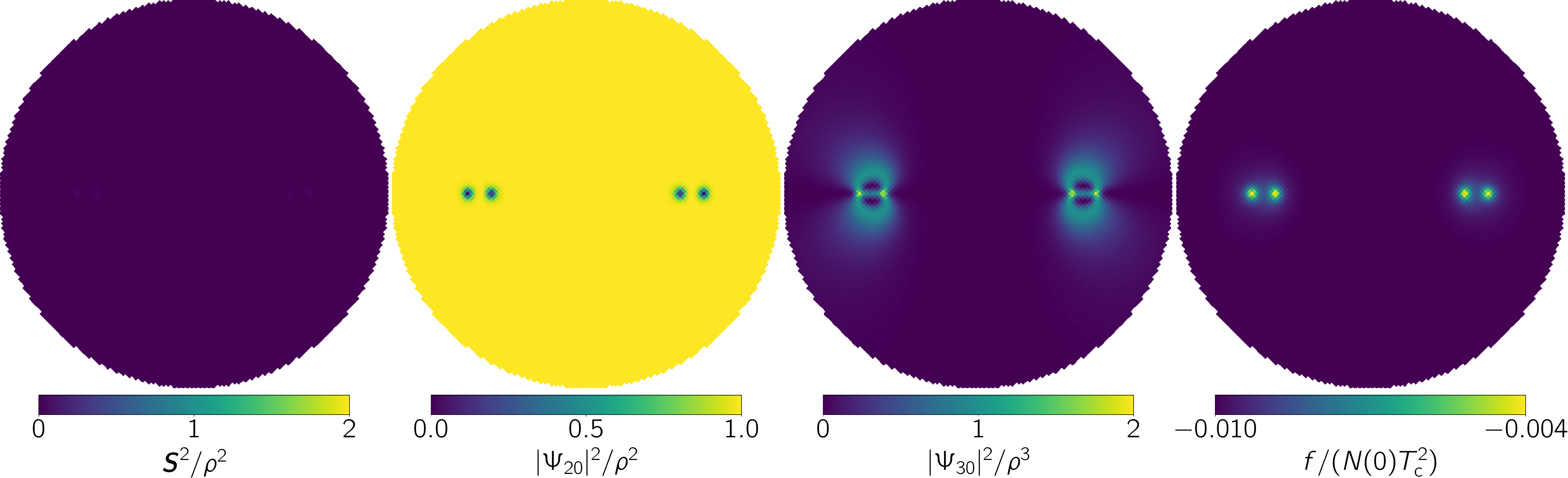}
    \caption{
        \label{fig:vortex-85-130-02-04}
        The two-vortex molecule state in the $D_4$BN phase at $\Vec{B}=1.3B_{\rm c}\hat{\Vec{z}}$ and $\Omega^2=0.4\Omega_0^2$.
        The squared modulus $|\psi_m|^2$ (top row), argument $\mathrm{Arg}[\psi_m]$ (middle row) of the order parameter, the $\mathrm{U}(1) \times \mathrm{SO}(3)$ invariants $\Vec{S}^2$, $|\Psi_{20}|^2$, and $|\Psi_{30}|^2$, and the free-energy density $f$ (bottom row) are shown. 
        The radius of the figure shown here is $64 p_{\rm F} / (\pi m_{\rm n} T_{\rm c}) \approx 9.29$ pm.
    }
\end{figure*}
Finally, for the $|\Vec{B}|\geq B_{\rm c}$ case in which the $D_4$BN phase is the ground state, we expect the C-core vortex molecule connected by three $D_2$BN solitons as shown in Fig.~\ref{fig:onevortex}(c).
Figure~\ref{fig:vortex-85-130-02-04} shows the two-vortex molecule state at $\Vec{B}=1.3B_{\rm c}\hat{\Vec{z}}$ and $\Omega^2=0.4\Omega_0^2$.
There are two isolated vortex molecules 
in each of which 
the three $D_2$BN solitons connect two half-quantized vortices.
At the boundary, the order parameter satisfies
\begin{align}
\psi |_{r=L}=\frac{e^{2i\theta}}{\sqrt{\rho}} \left(\frac{e^{-2 i a}}{\sqrt{2}},0,0,0,\frac{e^{2 i a}}{\sqrt{2}}\right)^T,
\label{eq:D4BN-double}
\end{align}
which is double winding of Eq.~\eqref{eq:D4BN-single}.
We define the molecule angle $\theta_{\rm molecule}$ as the angle between the polarization direction of each vortex molecule and 
the direction of separation of the centers of the two vortex molecules.
In Fig.~\ref{fig:vortex-85-130-02-04}, the molecule angle takes $\theta_{\rm rel}=0$.
On the other hand, for the $|\Vec{B}|<B_{\rm c}$ cases shown in 
Figs.~\ref{fig:vortex-85-000-02-08}, \ref{fig:vortex-85-000-02-10}, and \ref{fig:vortex-85-050-02-08}, the molecule angle takes $\theta_{\rm molecule}=\pi/2$.

\begin{figure}[htb]
    \centering
    \includegraphics[width=0.99\linewidth]{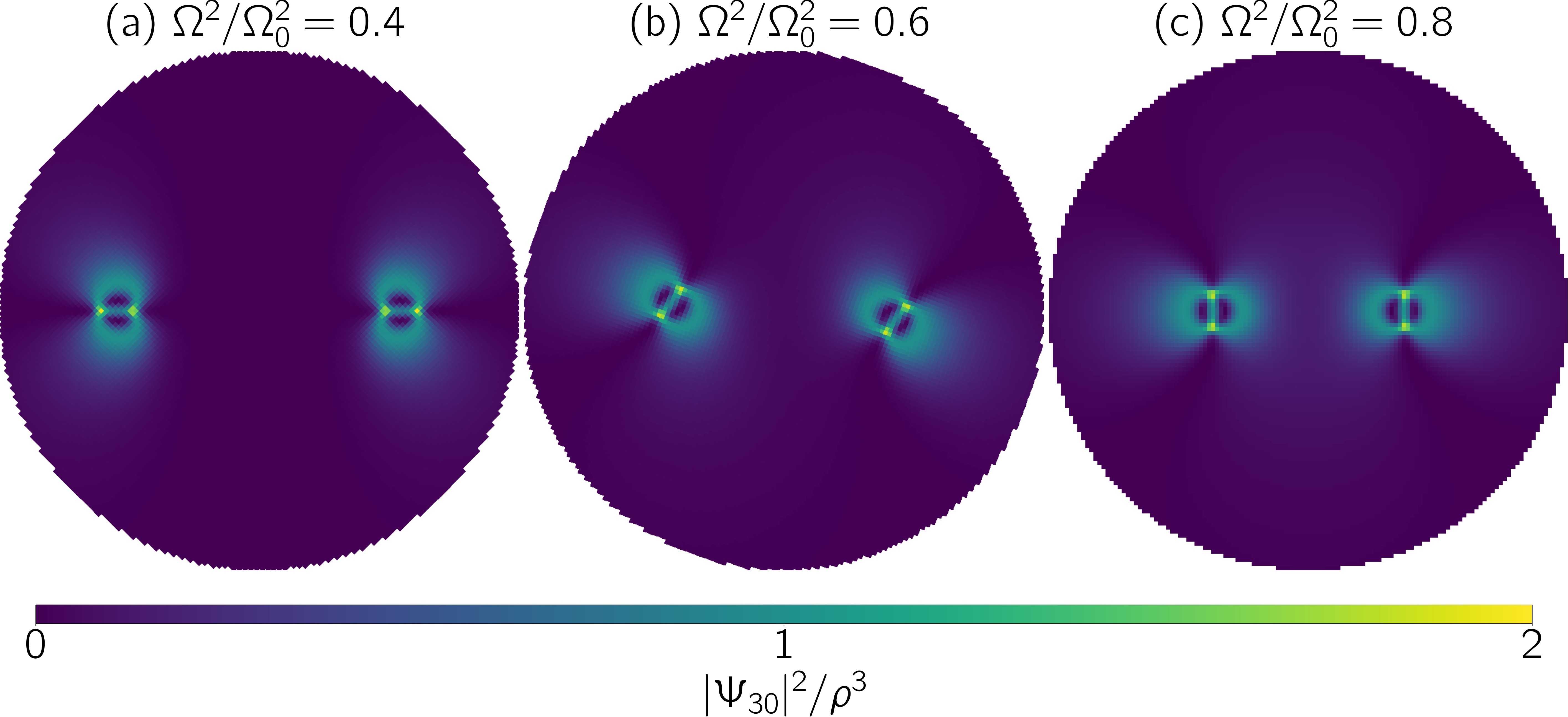}
    \caption{
    A transition of the polarization direction of the two-vortex molecules
    in the $D_4$BN phase. 
        \label{fig:vortices-85-130-02}
        $|\Psi_{30}|^2$ for two-vortex state in the $D_4$BN phase at $\Vec{B}=1.3B_{\rm c}\hat{\Vec{z}}$ and (a) $\Omega^2=0.4\Omega_0^2$, (b) $\Omega^2=0.6\Omega_0^2$, and (c) $\Omega^2=0.8\Omega_0^2$.
        The radius of the figure shown here is $64 p_{\rm F} / (\pi m_{\rm n} T_{\rm c}) \approx 9.29$ pm.
    }
\end{figure}
Being different from the $|\Vec{B}|<B_{\rm c}$ cases, the dimerization of vortex molecules never occurs even with increasing the rotation frequency $\Omega$.
On the other hand, a rapid change of $\theta_{\rm molecule}$ from $0$ to $\pi/2$ occurs.
Figure~\ref{fig:vortices-85-130-02} shows $|\Psi_{30}|^2$ profiles for the two-vortex molecule states at the rotation frequencies $\Omega^2/\Omega_0^2=0.4$ [panel (a)], $0.6$ [panel (b)], and $0.8$ [panel (c)], and the molecule angle $\theta_{\rm molecule}$ takes $\theta_{\rm molecule}\approx 0$, $0.22\pi$, and $\pi/2$, respectively. 

\subsubsection{Phase diagram}

\begin{figure}[htb]
    \centering
    \includegraphics[width=0.7\linewidth]{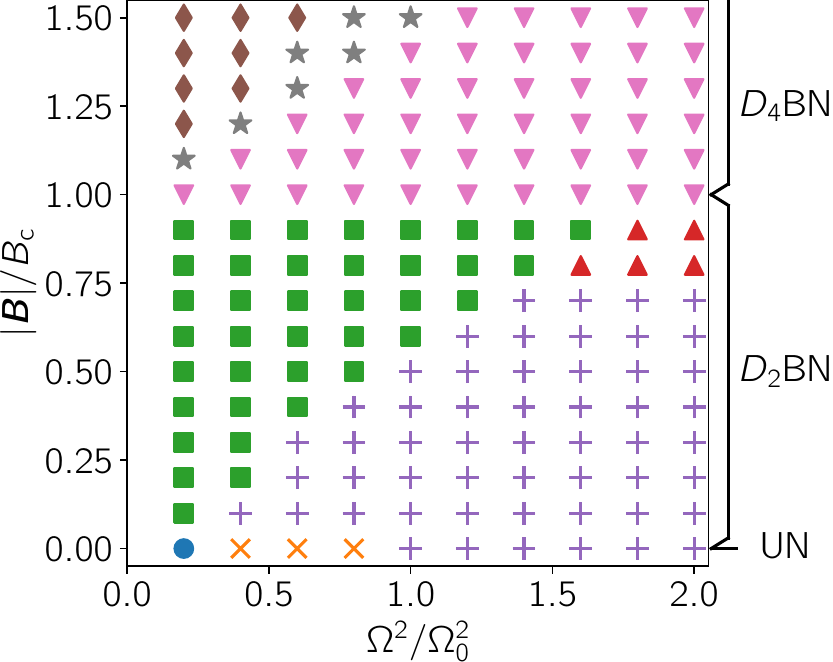}
    \caption{
        \label{fig:phase_diagram}
        The $\Omega$--$|\Vec{B}|$ phase diagram for two-vortex molecule states.
        \ding{108}: Coexistence of isolated stable F-core molecules and isolated metastable C-core molecules.
        $\times$: Coexistence of isolated stable F-core molecules, isolated metastable C-core molecules, dimerized stable molecule, and dimerized metastable molecule.
        $-$: Coexistence of isolated metastable C-core molecules, dimerized stable molecule, and dimerized metastable molecule.
        $\blacksquare$: isolated C-core molecules.
        $+$: Coexistence of isolated C-core molecules and dimerized molecule.
        $\blacktriangle$: Dimerized molecule.
        $\blacklozenge$: Molecules with $\theta_{\rm molecule}=0$.
        $\blacktriangledown$: Molecules with $\theta_{\rm molecule}=\pi/2$.
        $\bigstar$: Molecules with $0<\theta_{\rm molecule}<\pi/2$.
    }
\end{figure}
Figure~\ref{fig:phase_diagram} shows the $\Omega$-$|\Vec{B}|$ phase diagram for two-vortex molecule states.
The dimerized molecules for $|\Vec{B}|<B_{\rm c}$ and molecules with $\theta_{\rm molecule}=\pi/2$ can exist at large $\Omega$ and small $|\Vec{B}|$.
This result supports that these states originate from a proximity effect between two vortex molecules, because the distance between molecules decreases with increasing $\Omega$ and the size of a molecule decreases with increasing $|\Vec{B}|$ (see Fig.~5(b) in Ref.~\cite{Kobayashi:2022moc}).
At $|\Vec{B}|<B_{\rm c}$, the dimerized molecule state coexists with the isolated molecule state in a wide region of the phase diagram ($\times$, $-$, and $+$ symbols).
Here, the coexistence of several states means that one state is stable and others are metastable, or that some states are degenerate and equally stable.
\begin{figure}[htb]
    \centering
    \includegraphics[width=0.75\linewidth]{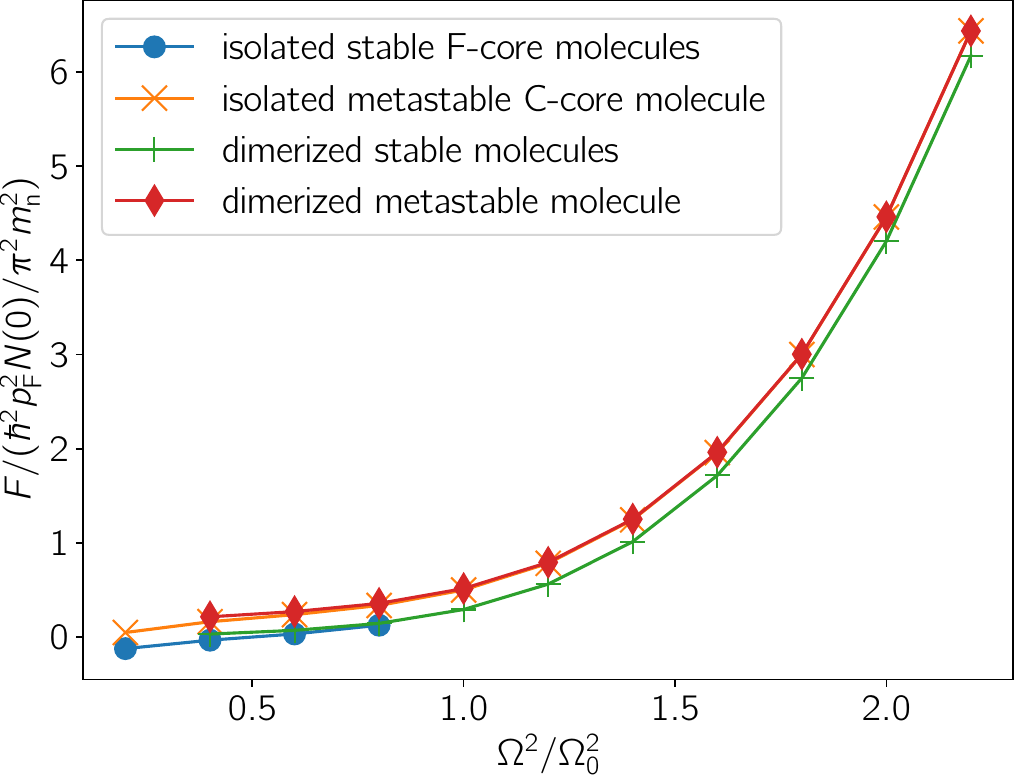}
    \caption{
        \label{fig:energy}
        Two-dimensional free energy $F = \int d^2 x\: f$ at $|\Vec{B}|=0$ as a function of $\Omega^2$ for the isolated stable F-core molecule state, the isolated metastable C-core molecule state, the dimerized stable molecule state, and the dimerized metastable molecule state.
    }
\end{figure}

To check this, we calculate the two-dimensional free energy $F=\int d^2x\: f$.
Figure~\ref{fig:energy} shows the two-dimensional free energy $F$ as a function of $\Omega^2$ at $|\Vec{B}|=0$ where the isolated stable F-core molecule state (see Fig.~\ref{fig:vortex-85-000-02-08}), the isolated metastable C-core molecule state (see Fig.~\ref{fig:vortex-85-000-02-08-ms}), the dimerized stable molecule state (see Fig.~\ref{fig:vortex-85-000-02-10}), and the dimerized metastable molecule state (see Fig.~\ref{fig:vortex-85-000-02-10-ms}) are stabilized.
The isolated stable F-core molecule state and dimerized stable molecule state
are stable states at small and large rotation frequencies $\Omega$, respectively,
having lower free-energy
than those of the isolated metastable C-core molecule state and the dimerized metastable molecule state.
On the other hand, the isolated stable F-core (metastable C-core) molecule state and the dimerized stable (metastable) molecule state have almost degenerate energies in a wide range of $\Omega$, which does not change in the case of finite values of $|\Vec{B}|$ ($+$ symbols in Fig.~\ref{fig:phase_diagram}).

\subsection{Three and four-vortex molecule states}

For more than two vortex molecule states, we expect various stable configurations such as completely isolated molecules, partially polymerized molecules, fully polymerized molecule at $|\Vec{B}|<B_{\rm c}$, and clusters of molecule pairs with various $\theta_{\rm molecule}$ at $|\Vec{B}|>B_{\rm c}$.
Actually, it is quite difficult to exhaust all (meta)stable states.

\begin{figure}[htb]
    \centering
    \includegraphics[width=0.99\linewidth]{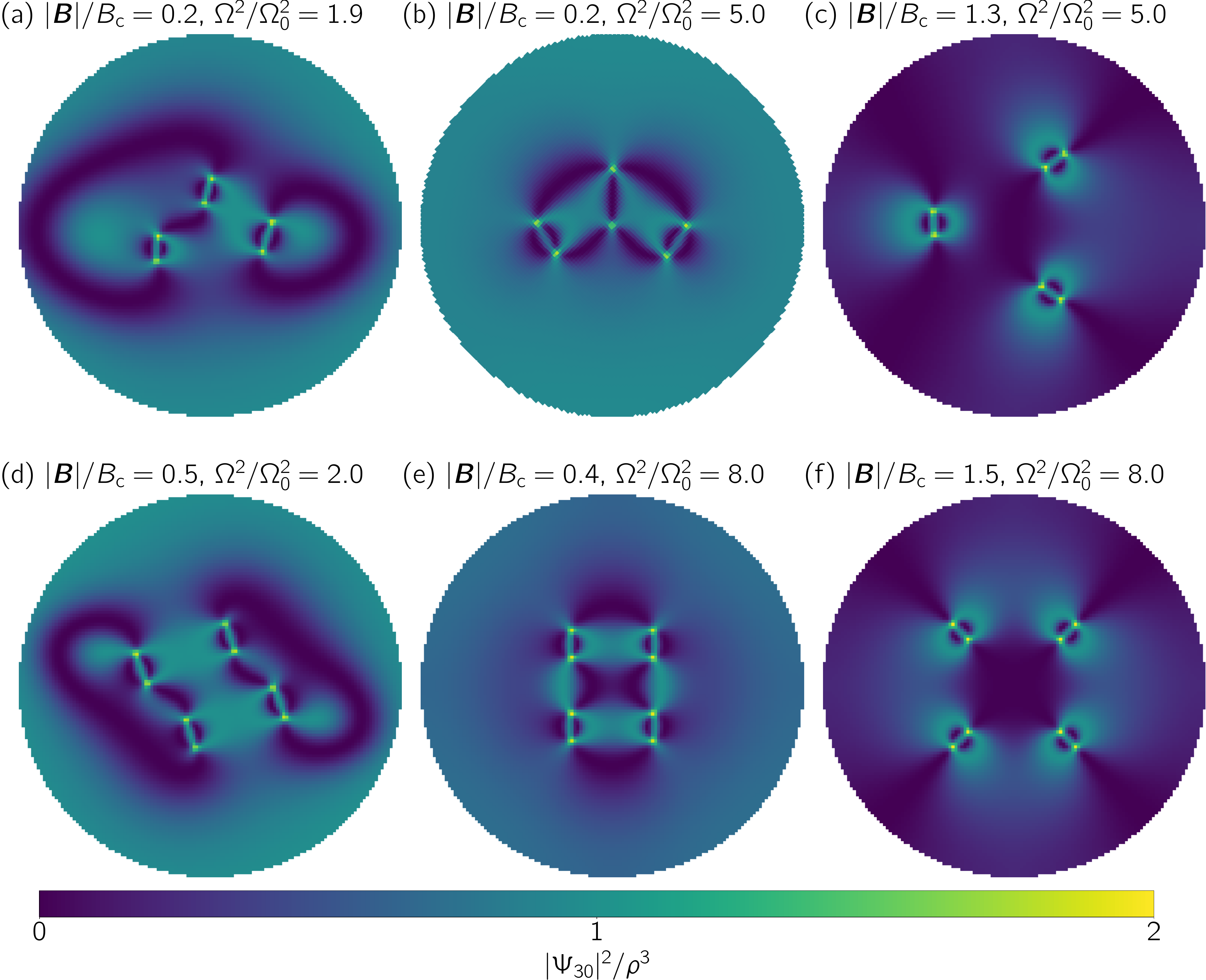}
    \caption{
        \label{fig:multi_vortex}
        $|\Psi_{30}|^2$ for three-vortex states (upper panels) and four-vortex states (lower panels) at
        (a) $\Vec{B}=0.2B_{\rm c}\hat{\Vec{z}}$ and $\Omega^2=1.9\Omega_0^2$,
        (b) $\Vec{B}=0.2B_{\rm c}\hat{\Vec{z}}$ and $\Omega^2=5.0\Omega_0^2$,
        (c) $\Vec{B}=1.3B_{\rm c}\hat{\Vec{z}}$ and $\Omega^2=5.0\Omega_0^2$,
        (d) $\Vec{B}=0.5B_{\rm c}\hat{\Vec{z}}$ and $\Omega^2=2.0\Omega_0^2$, and
        (e) $\Vec{B}=0.4B_{\rm c}\hat{\Vec{z}}$ and $\Omega^2=8.0\Omega_0^2$, and
        (f) $\Vec{B}=1.5B_{\rm c}\hat{\Vec{z}}$ and $\Omega^2=8.0\Omega_0^2$.
        The radius of the figure shown here is $64 p_{\rm F} / (\pi m_{\rm n} T_{\rm c}) \approx 9.29$ pm.
    }
\end{figure}
Instead, here we just show several characteristic examples for three and four-vortex states.
Figure~\ref{fig:multi_vortex} shows six examples of three and four vortex molecule states.
At $|\Vec{B}|<B_{\rm c}$, we have various kinds of ``polymerization'' of vortex molecules.
In Fig.~\ref{fig:multi_vortex}(a) and \ref{fig:multi_vortex}(d), we have one dimerized molecule and one isolated molecule, and two dimerized molecules, as three and four vortex molecule states, respectively.
We further obtain ``trimerized'' and ``tetramerized'' molecules 
in Figs.~\ref{fig:multi_vortex}(b) and \ref{fig:multi_vortex}(d), respectively.
At $|\Vec{B}|\geq B_{\rm c}$, there are also various kinds of clusters of vortex molecules.
We show symmetric examples of them in Figs.~\ref{fig:multi_vortex}(c) and \ref{fig:multi_vortex}(f).

\subsection{States with many vortex molecules}
\begin{figure}[htb]
    \centering
    \includegraphics[width=0.99\linewidth]{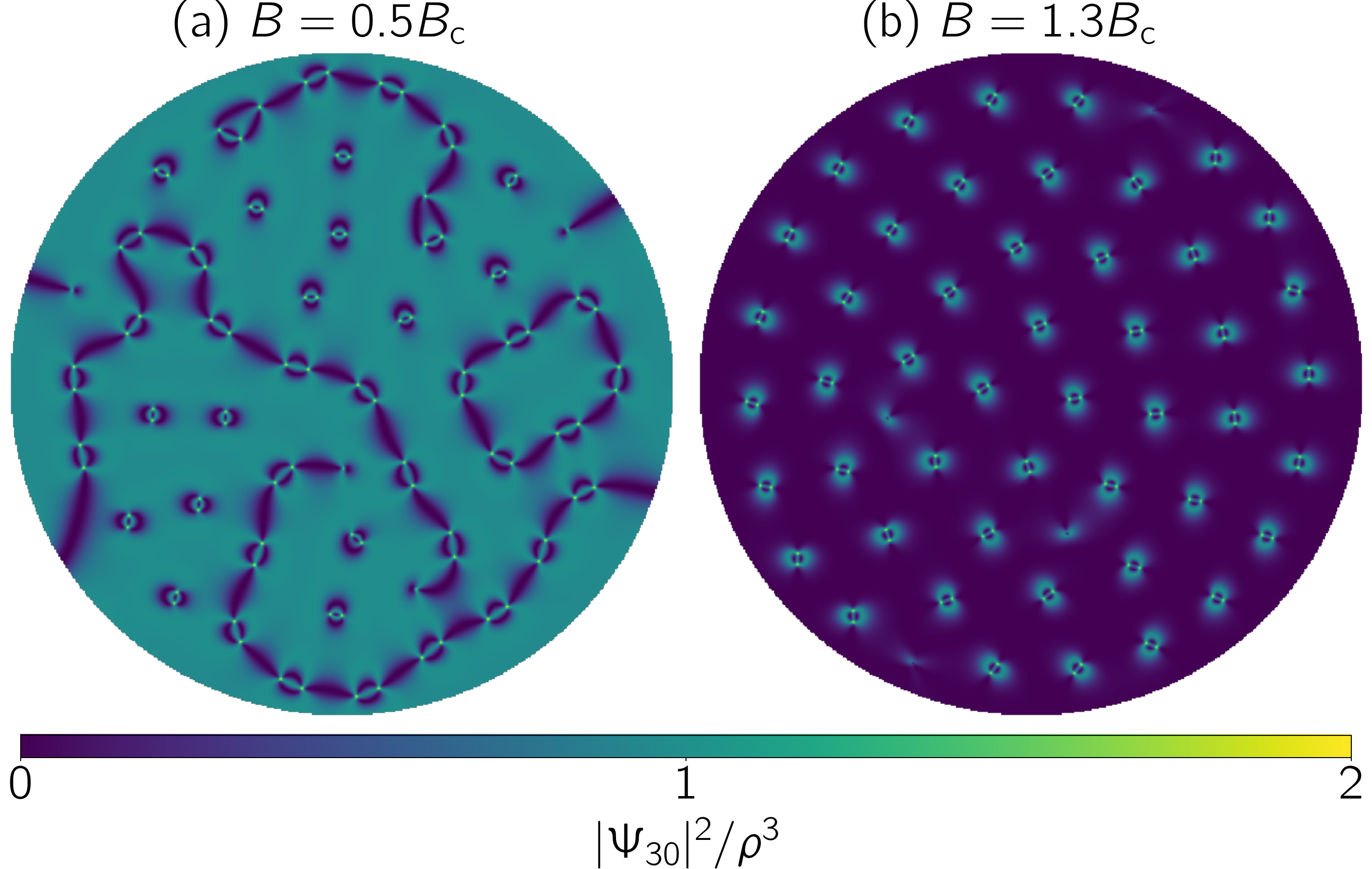}
    \caption{
        \label{fig:vortex_lattice}
        $|\Psi_{30}|^2$ for states with 50 vortices at
        (a) $\Vec{B}=0.5B_{\rm c}\hat{\Vec{z}}$ and $\Omega^2=\Omega_0^2$,
        (c) $\Vec{B}=1.3B_{\rm c}\hat{\Vec{z}}$ and $\Omega^2=\Omega_0^2$.
        The radius of the figure shown here is $256 p_{\rm F} / (\pi m_{\rm n} T_{\rm c}) \approx 37.2$ pm.
    }
\end{figure}
In the end of the section, we discuss a system with many vortices that can be candidates realized in rotating neutron star interiors.
Figure \ref{fig:vortex_lattice} shows systems having 50 vortices in the radius $256 p_{\rm F} / (\pi m_{\rm n} T_{\rm c}) \approx 37.2$ pm under the rotation $\Omega=\Omega_0$.
In panel (a) for $|\Vec{B}|<B_{\rm c}$, we obtain a mixture of single vortex molecules and long vortex chains polymerized through covalent bonds between vortex molecules.
In panel (b) for $|\Vec{B}|>B_{\rm c}$, on the other hand, vortex molecules form a lattice structure.
However, its structure is irregular and different from the regular triangular configuration that is expected for the vortex lattice in singlet-pairing superfluids.
We expect that this irregularity comes from a frustration between the spatial arrangement and internal alignment of vortex molecules.
As a consequence, we obtain irregular spatial structures of vortex molecules at both $|\Vec{B}|<B_{\rm c}$ and $|\Vec{B}|>B_{\rm c}$.

\section{Summary and Discussion}
\label{sec:summary} 

In this paper, we have worked out 
solutions for two vortex molecules consisting of four half-quantized non-Abelian vortices 
in the neutron $^3P_2$ superfluids in the presence of 
the external magnetic field 
parallel to the angular momentum of the vortices. 
We have found two characteristic transitions at $|\Vec{B}|<B_{\rm c}$ and $|\Vec{B}|\geq B_{\rm c}$ as a result of the proximity effect between the two vortex molecules.
At $0<|\Vec{B}|<B_{\rm c}$ where the C-core vortex molecule with three $D_4$BN solitons is stabilized in the $D_2$BN ground state, two isolated vortex molecules dimerize through covalent bonds of $D_4$BN solitons namely the solitons shared by the two molecules.
At $\Vec{B}=0$ where the F-core vortex molecule with a $D_4$BN soliton is the most stable state in the UN ground state, not only dimerization but also the transition of the vortex core state from the F state to 
the C state simultaneously occurs.
There are also metastable isolated C-core vortex molecules and the dimerized vortex molecule which are continuously changed from states at $0<|\Vec{B}|<B_{\rm c}$.
Our results suggest that a dimerized F-core vortex molecule never exist.
This situation is similar to the chemical dimerization in which two molecules having only a $\sigma$ bond cannot dimerize and those having more than two bonds consisting of a $\sigma$ bond and one or two $\pi$ bonds can dimerize.
Because the F-core vortex molecule has only one $D_4$BN soliton, the vortex core needs to change from F state to C state with three $D_4$BN solitons for isolated F-core vortex molecules to dimerize.
At $|\Vec{B}|\geq B_{\rm c}$ where the C-core vortex molecule with three $D_2$BN soliton is stabilized in the $D_4$BN ground state, 
no dimerization 
occurs.
Instead, a transition of the molecule angle from $\theta_{\rm molecule}=0$ to $\theta_{\rm molecule}=\pi/2$ occurs.
The main results are summarized as the $|\Vec{B}|$--$\Omega$ phase diagram shown in Fig.~\ref{fig:phase_diagram}.
We have also obtained some trimerized and tetramerized vortex molecule states as three and four vortex molecule solutions, respectively.
As candidates for what happens at neutron star interiors, solutions with many vortices have also been calculated.
Being different from regular triangular structure expected for singlet-pairing superfluids, the vortex configuration becomes irregular due to polymerizations of vortex molecules for $|\Vec{B}|<B_{\rm c}$ and a frustration between the spatial arrangement and the internal alignments of vortex molecules for $|\Vec{B}|>B_{\rm c}$.

Here, let us give discussions for future studies.
We have exhausted possible two-vortex molecule states, 
and have briefly given some examples of three, four, and many
vortex molecule states 
since these states are quite nontrivial.
Further detailed and systematic studies should be done 
for a vortex-molecule lattice under a rapid rotation 
relevant for neutron star interiors.
Although we show the nonsymmetric configurations of vortex molecules in this work,
detailed analyses are needed including comparisons with a symmetric vortex-molecule lattice as found in two-component BECs \cite{Mueller2002,Kasamatsu2003,Cipriani:2013nya}.
We here have studied only the case for the magnetic field parallel to the direction of vortices. 
It will be one of important directions 
for neutron star interiors 
to investigate the case of an arbitrary angle between them.

Apart from steady states under rotation studied thus far, 
a dynamics of vortices in three spatial dimensions 
is also crucial for neutron star dynamics 
such as pulsar glitch phenomena.
The most important feature of vortex dynamics in 
$^3P_2$ superfluids would be  
the non-Abelian property of half-quantized vortices 
implying that they are noncommutative under exchange.
It is important whether two vortices reconnect in collision or a formation of a rung between them occurs   
as the cases of non-Abelian vortices in the cyclic phase of a spin-2 BEC \cite{Kobayashi:2008pk} 
and the nematic phase of a spin-2 BEC \cite{Kobayashi:2011_arXiv,Borgh:2016cco}. 
For instance, a vortex reconnection 
is crucial for states of quantum turbulences \cite{Kobayashi:2016_arXiv}. 
Collision of two vortex molecules may be accompanied 
by swapping partners as in vortex molecules in two-component BECs \cite{Eto:2019uhe}. 
In addition,
the dimerization of vortex molecules found in this paper 
may occur in a certain part of two vortex lines, 
thus forming for instance an X junction.
Thus, multiple vortex molecules are expected to be 
entangled in general, 
which should be also crucial for 
pulsar glitches.

Let us mention the case of two spatial dimensions.
A phase transition due to  
unbinding of a vortex and an anti-vortex 
is known as the Berezinskii-Kosterlitz-Thouless (BKT) transition 
in two spatial dimension.
Recently, a novel type of BKT transition due to   
vortex molecules
was found  in two-component systems \cite{Kobayashi:2018ezm}.
Therefore, the BKT transition of neutron $^3P_2$ superfluids 
confined in a quasi-two-dimensional plane 
should be an interesting subject. 

Here, let us discuss other phases of 
$^3P_2$ superfluids. 
In this paper, we have not considered a $^1S_0$ paring, but 
the phase diagram under 
the coexistence of $^1S_0$ and $^3P_2$ superfluids 
is quite different from the one of solely $^3P_2$ superfluids 
and the most region is occupied by the $D_4$BN phase 
\cite{Yasui:2020xqb}.
Vortex states in this case will be also one direction to be 
explored. 
In addition, 
vortex states in the ferromagnetic phase, 
that was found  in the region 
close to the critical temperature 
without quasiclassical approximation
\cite{Mizushima:2021qrz},
are also worth to be studied. 

Quark matter consisting of diquark condensations 
exhibiting color superconductivity may exist 
in the region deeper than nuclear matter in neutron star cores. 
A quark-hadron continuity for two-flavor quarks 
was suggested to 
continuously connect 
the $^3P_2$ superfluid (nuclear matter) to
a two-flavor quark matter called the 2SC + dd phase 
through crossover (rather than a phase transition)
\cite{Fujimoto:2019sxg}, 
and 
vortex structures in the 2SC + dd phase were studied in 
Refs.~\cite{Fujimoto:2020dsa,Fujimoto:2021bes,Fujimoto:2021wsr}.
It will be interesting to study whether vortex molecule structures 
found in this paper is preserved or deformed through 
the quark-hadron continuity.

Finally, 
apart from $^3P_2$ superfluids, 
spin-2 spinor ultracold atomic BECs  
are also $J=2$ condensates 
whose ground states are possibly nematic phase  
\cite{Zhou:2006fd,Semenoff:2006vv,
Uchino:2010,Uchino:2010pf,Borgh:2016cco,Kobayashi:2021arv} 
although the current experiments of $^{87}$Rb atoms imply their 
ground state to be in the cyclic or nematic phase 
\cite{Schmaljohann:2004,Chang:2004,Kuwamoto:2004,Widera:2006,Tojo:2008,Tojo:2009}. 
Nematic spin-2 BECs
share almost the same bosonic properties with $^3P_2$ superfluids, 
and thus admit the same order parameter manifold and 
non-Abelian half-quantized vortices 
\cite{Uchino:2010pf,Borgh:2016cco}.
Therefore, our present results for the dimerization of vortex molecules are also applicable to spin-2 nematic BECs 
which can be experimentally testable in principle.

\section*{Acknowledgments}

We would like to thank Yusuke Masaki for helpful discussions and comments.
This work is supported in part by 
 JSPS KAKENHI [Grants No. JP22H01221 (M.K. and M.N.),
 20K03765(M.K.), 19KK0066(M.K.)], the WPI program ``Sustainability with Knotted Chiral Meta Matter (SKCM$^2$)'' at Hiroshima University, and by Osaka Metropolitan University Advanced Mathematical Institute (MEXT Joint Usage/Research Center on Mathematics and Theoretical Physics JPMXP0619217849).

\newpage

\bibliography{vortex_only}

\end{document}